\documentclass[pdflatex]{sn-jnl} 

\usepackage[T1]{fontenc}
\usepackage[utf8]{inputenc}
\usepackage{csquotes}

\usepackage{amsfonts}       
\usepackage{amsmath}
\usepackage{amsthm}%
\usepackage{siunitx}
\usepackage{mathtools}
\usepackage{braket}
\usepackage{pifont}
\usepackage{nicefrac}       

\usepackage{aligned-overset}
\usepackage{geometry}
\usepackage{graphicx, import}
\usepackage{tikz}
\usepackage[compat=1.1.0]{tikz-feynman}
\tikzfeynmanset{warn luatex=false} 
\usetikzlibrary{quantikz2}
\usepackage{tabularx}
\usepackage{booktabs}

\usepackage{parskip}
\usepackage{caption}
\usepackage{subcaption} 
\usepackage[title]{appendix}%
 
\usepackage{xcolor}
\usepackage{orcidlink}
\usepackage{soul}
\usepackage{hyperref}
\usepackage{url}            
\hypersetup{colorlinks=true,urlcolor=teal,
            linkcolor=teal,citecolor=teal} 
\usepackage{glossaries-extra}

\usepackage[export]{adjustbox}
\usepackage{cleveref}       
\usepackage{csquotes}
\usepackage{xspace}

\usepackage{microtype}      

\usepackage{stmaryrd} 
\SetSymbolFont{stmry}{bold}{U}{stmry}{m}{n} 

\usepackage{silence}
\usepackage[
  backend=biber,
  maxcitenames=3,
  minbibnames=3,
  mincitenames=1,
  maxbibnames=3,
  uniquelist=false,
  url=false,
  doi=true,
  eprint=false,
  isbn=false,
]{biblatex}
\newcommand{\eg}{\emph{e.g.}\xspace}
\newcommand{\ie}{\emph{i.e.}\xspace}
\newcommand{\cf}{\emph{c.f.}\xspace}

\newcommand{\ketn}[1]{\vert#1\rangle^{\otimes{n}}}
\newcommand{\bran}[1]{\langle#1\vert^{\otimes{n}}}

\newcommand{\Ase}{Ansatzes\xspace}
\newcommand{\ase}{ansatzes\xspace}

\newcommand{\as}{ansatz\xspace}

\newcommand{\btheta}{\ensuremath \boldsymbol{\theta}}
\newcommand{\bomega}{\ensuremath \boldsymbol{\omega}}

\newcommand{\bOmega}{\ensuremath \boldsymbol{\Omega}}
\newcommand{\poly}{\ensuremath \mathrm{poly}}
\newcommand{\bx}{\ensuremath \boldsymbol{x}}

\newcommand{\sizex}{\ensuremath \vert \mathcal{X} \vert}
\newcommand{\sizefreq}{\ensuremath \vert \bOmega \vert}
\newcommand{\sizeparams}{\ensuremath \vert \btheta \vert}
\newcommand{\fullspec}{\ensuremath \tilde{\bOmega}}
\newcommand{\uniquespec}{\ensuremath \bOmega}
\newcommand{\fcc}{\ensuremath \mathtt{FCC}_{\Theta}}
\newcommand{\fccw}{\ensuremath \overline{\mathtt{FCC}}_{\Theta}}
\newcommand{\hnscrit}{\ensuremath \max(\Omega)}
\newcommand{\nscrit}{\ensuremath 2\hnscrit}
\newcommand{\exptheta}{\ensuremath \mathbb{E}_{\Theta}}
\newcommand{\loss}{\ensuremath \mathcal{L}}
\newcommand{\var}{\ensuremath \text{Var}}
\newcommand{\imag}{\mathrm{i}}

\definecolor{color1}{HTML}{009682}
\definecolor{color2}{HTML}{DF9B1B}
\definecolor{color3}{HTML}{23A1E0}

\theoremstyle{thmstyleone}%
\theoremstyle{thmstyletwo}%

\theoremstyle{thmstylethree}%

\begin{document}
\setabbreviationstyle[acronym]{long-short}
\glssetcategoryattribute{acronym}{nohyperfirst}{true}
\renewcommand*{\glstextformat}[1]{\textcolor{black}{#1}}

\renewcommand*{\glsdonohyperlink}[2]{%
    {\glsxtrprotectlinks \glsdohypertarget{#1}{#2}}}

\newacronym{fcc}{FCC}{Fourier coefficient correlation}
\newacronym{fft}{FFT}{fast Fourier transform}
\newacronym{fm}{FM}{feature map}
\newacronym{ftqc}{FTQC}{fault‑tolerant quantum computing}
\newacronym{hea}{HEA}{hardware‑efficient ansatz}
\newacronym{hep}{HEP}{high‑energy physics}
\newacronym{kl}{KL}{Kullback‑Leibler}
\newacronym{lhc}{LHC}{Large Hadron Collider}
\newacronym{lo}{LO}{leading order}
\newacronym{ml}{ML}{machine learning}
\newacronym{mlp}{MLP}{multi‑layer perceptron}
\newacronym{mse}{MSE}{mean squared error}
\newacronym{nisq}{NISQ}{noisy intermediate‑scale quantum}
\newacronym{nlo}{NLO}{next‑to‑leading order}
\newacronym{pca}{PCA}{principal component analysis}
\newacronym{pcc}{PCC}{pearson correlation coefficient}
\newacronym{pqc}{PQC}{parameterized quantum circuit}
\newacronym{qaoa}{QAOA}{quantum approximate optimisation algorithm}
\newacronym{qc}{QC}{quantum computing}
\newacronym{qcd}{QCD}{quantum chromodynamics}
\newacronym{qfm}{QFM}{quantum Fourier model}
\newacronym{qk}{QK}{quantum kernel}
\newacronym{qml}{QML}{quantum machine learning}
\newacronym{qnn}{QNN}{quantum neural network}
\newacronym{rff}{RFF}{random Fourier features}
\newacronym{spam}{SPAM}{state preparation and measurement}

\title[Fourier Fingerprints of Ansatzes in Quantum Machine Learning]{Fourier Fingerprints of Ansatzes \linebreak[1] in Quantum Machine Learning}

\author*[1]{\fnm{Melvin} \sur{Strobl} \orcidlink{0000-0003-0229-9897}}\email{melvin.strobl@kit.edu}
\author[2]{\fnm{M. Emre} \sur{Sahin} \orcidlink{0000-0002-5996-0407}}
\author[1]{\fnm{Lucas} \sur{van der Horst} \orcidlink{0009-0003-0609-9582}}
\author[1]{\linebreak[1]\fnm{Eileen} \sur{Kuehn} \orcidlink{0000-0002-8034-8837}}
\author[1]{\fnm{Achim} \sur{Streit} \orcidlink{0000-0002-5065-469X}}
\author[3]{and \fnm{Ben} \sur{Jaderberg} \orcidlink{0000-0001-9297-0175}}

\affil*[1]{%
    \orgname{Karlsruhe Institute of Technology},\linebreak[1]
    \orgaddress{76344 Eggenstein-Leopoldshafen, \country{Germany}}%
}

\affil[2]{%
    \orgname{The Hartree Centre, STFC},\linebreak[1]
    \orgaddress{Sci-Tech Daresbury, Warrington, WA4 4AD, \country{United Kingdom}}%
}

\affil[3]{%
    \orgname{IBM Quantum, IBM Research Europe},\linebreak[1]
    \orgaddress{Hursley, Winchester, SO21 2JN, \country{United Kingdom}}%
}

\abstract{
    \unboldmath
    Typical schemes to encode classical data in variational \gls{qml} lead to quantum Fourier models with $\mathcal{O}(\exp(n))$ Fourier basis functions in the number of qubits.
    Despite this, in order for the model to be efficiently trainable, the number of parameters must scale as $\mathcal{O}(\mathrm{poly}(n))$.
    This imbalance implies the existence of correlations between the Fourier modes, which depend on the structure of the circuit.
    In this work, we demonstrate that this phenomenon exists and show cases where these correlations can be used to predict \as performance.
    For several popular \ase, we numerically compute the \glspl{fcc} and construct the \emph{Fourier fingerprint}, a visual representation of the correlation structure.
    We subsequently show how, for the problem of learning random Fourier series, the \gls{fcc} correctly predicts relative performance of \ase whilst the widely-used expressibility metric does not.
    Finally, we demonstrate how our framework applies to the more challenging problem of jet reconstruction in \glsxtrlong{hep}.
    Overall, our results demonstrate how the Fourier fingerprint is a powerful new tool in the problem of optimal \as choice for \gls{qml}.
}

\keywords{Quantum Computing, Quantum Machine Learning, Quantum Fourier Models, Ansatz Selection, High-Energy Physics}

\maketitle

\glsresetall
\section{Introduction}

The field of \gls{qml} has emerged as one of the most rapidly advancing areas of quantum computing in recent years~\cite{cerezo_challenges_2022}.
In particular, \gls{qml} has seen a significant focus on the use of \glspl{pqc} as machine learning models~\cite{benedetti_parameterized_2019, mitarai_quantum_2018}.
Here, in a way that is conceptually similar to deep neural networks, the parameters of a quantum circuit are trained to learn the function $f(\bx, \btheta)$ that best approximates a given dataset of input-output pairs $(\bx, f(\bx, \btheta))$.
Due to its applicability to a wide range of domains~\cite{havlicek_supervised_2019, lloyd_quantum_2020, nagano_quantum_2023, jaderberg_potential_2024} and amenability to running on pre-fault-tolerant quantum computers~\cite{ pan_experimental_2023, chen_quantum_2025}, \glspl{pqc} remain one of the most widely-used paradigms in \gls{qml}.

In this framework, a typical \gls{qml} model consists of quantum \glspl{fm} $\hat{U}_F(\bx)$, which encode a classical input $\bx$ into the Hilbert space, and variational \ase $\hat{U}_A(\btheta)$ which manipulate the input via trainable parameterized gates $\btheta$.
Importantly, the output of these models can be represented as a partial Fourier series in the input~\cite{schuld_effect_2021}, giving rise to the name \glspl{qfm}.
Fully understanding how the choice of feature map and \as affects the Fourier spectrum of \glspl{qfm}, both analytically~\cite{wiedmann_fourier_2024} and numerically~\cite{jaderberg_let_2023,mhiri_constrained_2024, duffy_spectral_2025}, remains crucial for maximizing performance and probing dequantization~\cite{sweke_potential_2025}.

\begin{figure}[htb]
    \centering
    \includegraphics[width=\linewidth]{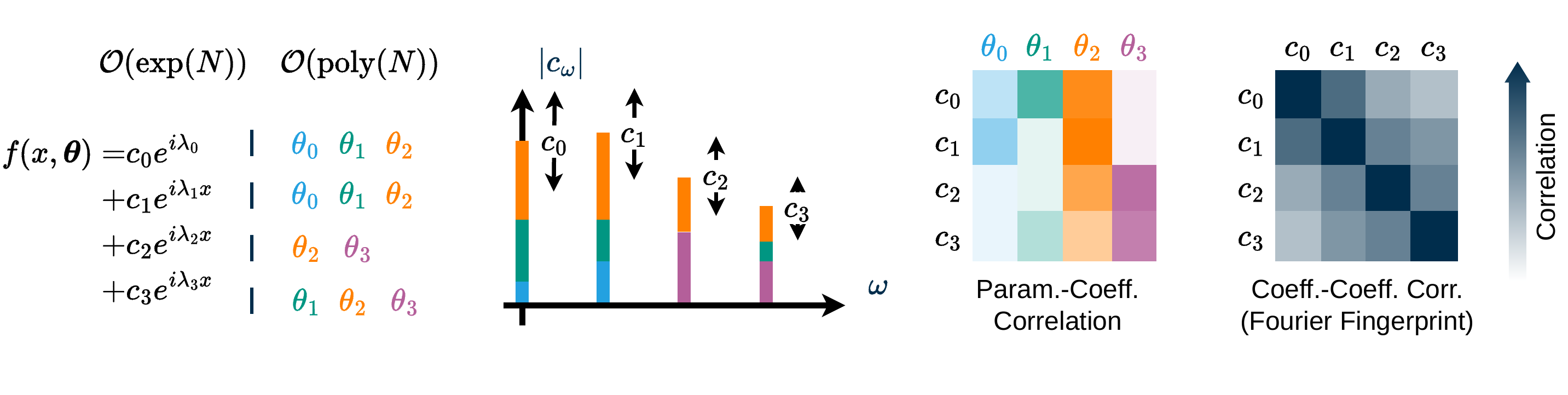}
    \caption{The output $f(x,\btheta)$ of an $n$ qubit \gls{qfm} is a truncated Fourier series with $\mathcal{O}(\exp(n))$ terms. To be efficiently trainable, this can only be tuned by $\mathcal{O}(\poly(n))$ parameters. This leads to a subset of the Fourier coefficients depending on the same parameters, inducing an effective correlation between them. Practically, this implies that \glspl{qfm} cannot independently control each frequency.}
    \label{fig:overview}
\end{figure}

In this work, we demonstrate a novel phenomenon in which correlations naturally arise between the coefficients of the Fourier series of efficiently-trainable \glspl{qfm}.
This implies that \glspl{qfm} are not able to control each term in their Fourier series independently, limiting the space of functions that can be practically learned.
We first theoretically motivate this in~\autoref{sec:method:correlations_in_qfm} and subsequently numerically validate the existence of these correlations for a range of popular \ase in~\autoref{sec:method:fourier_fingerprint}.
Furthermore, we find the specific correlation pattern to be unique for each \as, which we thus call the \emph{Fourier fingerprint} of the model.
An overview of this concept is pictured in~\autoref{fig:overview}.

Subsequently, we demonstrate in practice how the \emph{\gls{fcc}}, the average of the fingerprint, can be used to identify the best-performing \ase in~\autoref{sec:results:random_fourier_series}.
We find that, for learning random 1D Fourier series, models with lower \gls{fcc} perform better, even in cases where expressibility would predict the \as to be worse.
Furthermore, we extend our experiments to learning 2D Fourier series, where we observe the same result.

Finally, in~\autoref{sec:results:hep}, we show how the \gls{fcc} can be used to pick well-performing \ase in the more challenging problem of 2D jet reconstruction in high-energy physics.
Given the energy and momentum of a two-particle collision, we train our model to predict the transverse momentum of the largest jet.

Comparing the performance of different \ase, we find a similar trend as observed in the experiments concerning random Fourier series, \ie that models with lower average \gls{fcc} achieve lower \gls{mse}.
Overall, our results highlight that the Fourier fingerprint and the resultant \gls{fcc} metric are practically useful and should be integrated into the process of \as choice in \gls{qml}.

\section{Method}
\label{sec:method}

\subsection{Correlation between coefficients in quantum Fourier models}
\label{sec:method:correlations_in_qfm}

Let us define an $L$-layer \gls{qfm} by $U(\bx, \btheta) = W^{(L+1)}(\btheta)S(\bx)W^{(L)}\dots W^{(2)}S(\bx)W^{(1)}$, with trainable unitaries $W$ and $n$-qubit encoding gates $S$ taking a $D$ dimensional input $\bx$. For this class of models, which includes \glspl{qnn}, the output $f(\bx, \btheta)  =\bran{0} U^{\dagger}(\bx, \btheta) \mathcal{M} U(\bx, \btheta)\ketn{0}$ is obtained by measuring an expectation value of an observable $\mathcal{M}$.
The seminal work of Ref.~\cite{schuld_effect_2021} showed the equivalence of this expression to a truncated Fourier series
\begin{equation}
    f(\bx, \btheta) = \sum_{\bomega \in \uniquespec} c_{\bomega}(\btheta) \exp{(\imag \bomega \bx)}
    \label{eq:fourier_series_simple}
\end{equation}
where $\uniquespec$ defines the set of frequencies and $c_{\bomega}$ the complex valued coefficient belonging to the frequency vector $\bomega$.

From~\autoref{eq:fourier_series_simple}, we first note that the Fourier coefficients of the model are explicitly dependent on the trainable parameters $\btheta \in \Theta$ of the \as \footnote[1]{The coefficients also implicitly depend on the \gls{fm} as the degeneracy of the encoding spectrum upper-bounds the variance of the coefficients~\cite{mhiri_constrained_2024}.}.
The frequency spectrum of this \gls{qfm}, and specifically models where the output is an expectation value, is defined by the gaps in the eigenspectrum of the \gls{fm} generator Hamiltonian~\cite{schuld_effect_2021, mhiri_constrained_2024}
\begin{equation}
    f(\bx, \btheta) = \sum_{i=1}^{D} \sum_{\boldsymbol j, \boldsymbol k \in \llbracket 1, d_l \rrbracket^L_{l=1}} c_{\boldsymbol j, \boldsymbol k}(\btheta) \exp{(\imag (\Lambda_{i, \boldsymbol j} - \Lambda_{i, \boldsymbol k}) \boldsymbol e_i^T x)} \quad \mid \quad \boldsymbol{j}, \boldsymbol{k} \in \llbracket 1, d_l \rrbracket^L_{l=1},
    \label{eq:fourier_series}
\end{equation}
where $\llbracket \cdot \rrbracket$ defines the product of the intervals $j_l, k_l\in [1,d_l]$ in the multi-index $\boldsymbol{j}, \boldsymbol{k}$ and $\Lambda_{i, \boldsymbol j} = \lambda_{j_1}^i + \lambda_{j_2}^i + \ldots + \lambda_{j_L}^i$ is the sum of all $d_l$ eigenvalues of the $l$-th generator in $L$ layers for input index $i$. Whilst typically $\uniquespec$ is defined as the \emph{unique} gaps in the spectrum, here it is useful to define a larger set of frequencies that includes degenerate values
\begin{equation}
    \fullspec=\biguplus_{i=1}^{D} \left\{\Lambda_{i, \boldsymbol{j}}-\Lambda_{i, \boldsymbol{k}}\right\} \quad \mid \quad \boldsymbol{j}, \boldsymbol{k} \in \llbracket 1, d_l \rrbracket^L_{l=1}
    \label{eq:set_of_frequencies},
\end{equation}
Importantly, whilst the scaling of unique frequencies is determined by the degeneracy of the \gls{fm} eigenspectrum ~\cite{schuld_circuit_2020}, the absolute number of terms in the Fourier series always grows as $\vert\fullspec\vert \sim \mathcal{O}(2^{nD})$.

Finally, we note that for \glspl{qfm} to be efficiently trainable, they must contain a sub-exponential $\sizeparams \sim \mathcal{O}(\mathrm{poly}(n))$ number of parameters, since the time complexity of training is linear in $\sizeparams$. Combining these observations about the size of $\vert\fullspec\vert$ and $\sizeparams$, it is immediately apparent that in the asymptotic limit there is no injective map between the set of Fourier coefficients $c_\omega$ and $\btheta$.
Therefore, the coefficients $c_\omega(\btheta)$ must depend on a \emph{shared} set of \as parameters, which implies the existence of correlations between the coefficients of different terms in the Fourier series.
We provide another perspective on this statement in~\autoref{app:coeff_correlations}.

The practical implication of this result is that the training of \glspl{qfm} is in fact an optimization of Fourier series where each term is not independent. This is an additional constraint, which is not apparent from the form of~\autoref{eq:fourier_series_simple}, that reduces the possible solution space. Furthermore, this is in contrast to classical approaches to Fourier regression, where modes are tuned independently~\cite{brooks_fitting_2012, guruswami_robust_2016}.
In~\autoref{sec:results} we numerically show how this constraint for \glspl{qfm} affects performance, whilst in~\autoref{app:coeff_correlations_mse} we analytically consider how this constraint appears in the \gls{mse} loss.

\subsection{Computing the Fourier fingerprint}
\label{sec:method:fourier_fingerprint}

Given a \gls{qfm}, we numerically compute the correlations between Fourier coefficients as follows.
For a randomly initialized set of \as parameters $\btheta$, we evaluate $f(\bx, \btheta)$ for many different $\bx \in \mathcal{X}$ such that the number of samples satisfies the Shannon-Nyquist~\cite{shannon_communication_1949} criterion.
We then obtain the Fourier coefficients $\boldsymbol c$ using the \gls{fft}, specifically as implemented in the QML-Essentials framework~\cite{strobl_qml_2025}.
As the \gls{fft} requires knowledge of the number of frequencies in advance, in the remainder of this work we consider a single-qubit Pauli encoding, such that $\uniquespec = \llbracket -nL, nL \rrbracket^D$, based on~\autoref{eq:set_of_frequencies} under the assumption $d_l=2 \ \forall\  l\in[1\dots L]$.

We then repeat this procedure $M$ times for randomly sampled \as parameters $\btheta$ from the space of possible \as parameters $\Theta$, obtaining a set of $M$ Fourier coefficient vectors.
Specifically, each $\theta \in \btheta$ is i.i.d. and drawn from a uniform distribution with the interval $[0,2\pi]$.
Treating each calculated instance of $\boldsymbol c$ as independent samples, we then compute the Pearson correlation coefficient ~\cite{freedman_statistics_2007}
\begin{equation}
    r_{\Theta}(\bomega, \bomega')= \frac{\sum_{\btheta\in\Theta}\left(c_{\bomega}(\btheta)-\bar c_{\bomega}\right)\left(c_{\bomega'}(\btheta)-\bar c_{\bomega'}\right)}{\sqrt{\sum_{\btheta\in\Theta}\left(c_{\bomega}(\btheta)-\bar c_{\bomega}\right)^{2} \sum_{\btheta\in\Theta}\left(c_{\bomega'}(\btheta)-\bar c_{\bomega'}\right)^{2}}}
    \label{eq:pearson_correlation}
\end{equation}
between each possible pair in the spectrum.
Here, $\bar c_{\bomega}$ and $\bar c_{\bomega'}$ represent the average value of the coefficient at frequency $\bomega$ and $\bomega'$ respectively.
Note that the Pearson correlation is normalized by the standard deviation of the coefficients at each frequency.
We provide a sanity check of this in~\autoref{fig:coefficient_correlation_surrogate} where we randomly sample coefficients with an increasing number of redundancies towards higher frequencies.
Computing the correlation between them results in zeros for all frequency pairs, which confirms that the variance is mitigated, meaning that the decaying variance towards higher frequencies as observed in~\cite{mhiri_constrained_2024} is taken into account with the \gls{fcc} metric.

\begin{figure}[htb]
    \begin{subfigure}[b]{0.55\textwidth}
        \includegraphics[clip, trim=0.5cm 0.0cm 0.0cm 2.0cm, width=\textwidth]{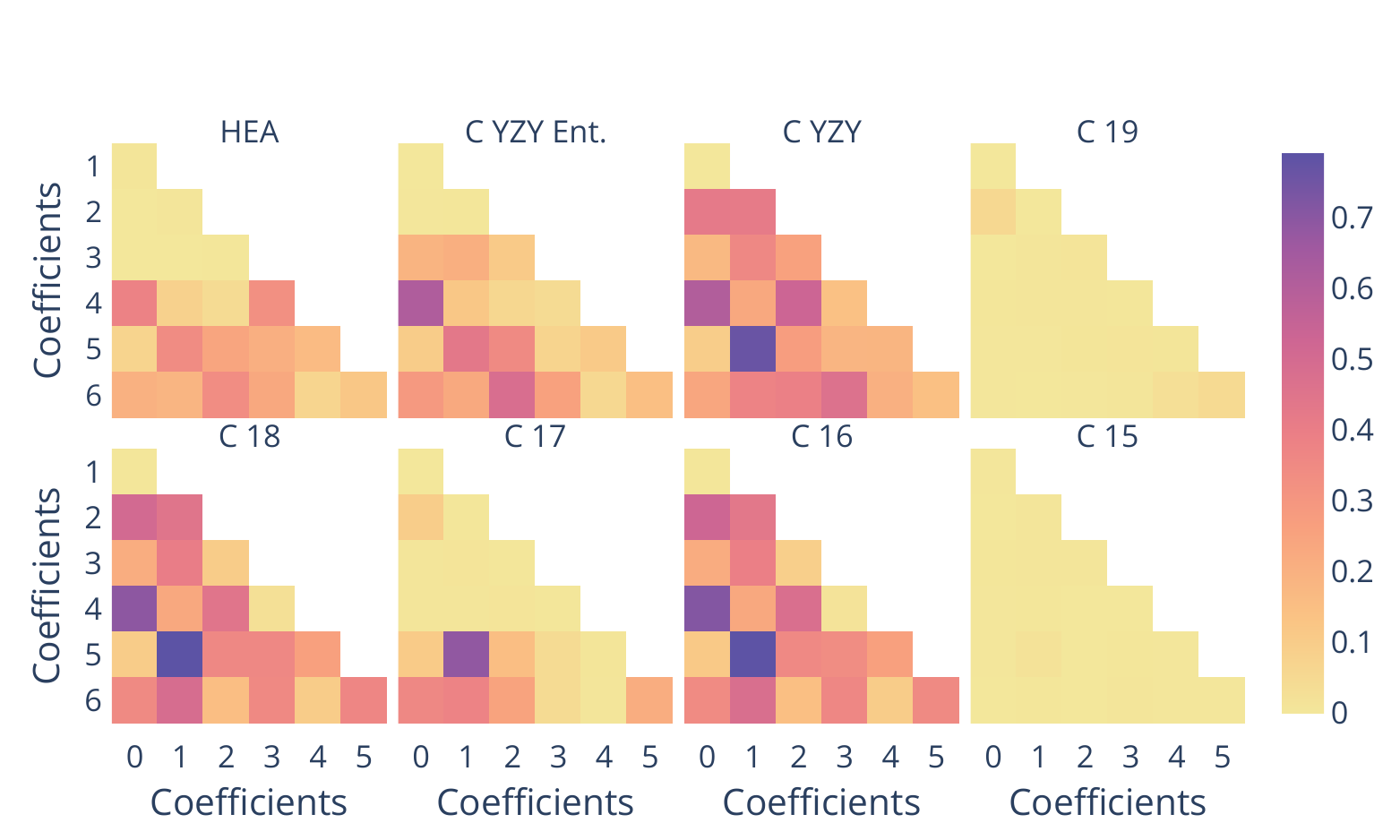}
        \caption{}
        \label{fig:fingerprints_1d_ry}
    \end{subfigure}\hfill
    \begin{subfigure}[b]{0.36\textwidth}
        \includegraphics[clip, trim=0.2cm 0.0cm 0.0cm 2.0cm, width=\textwidth]{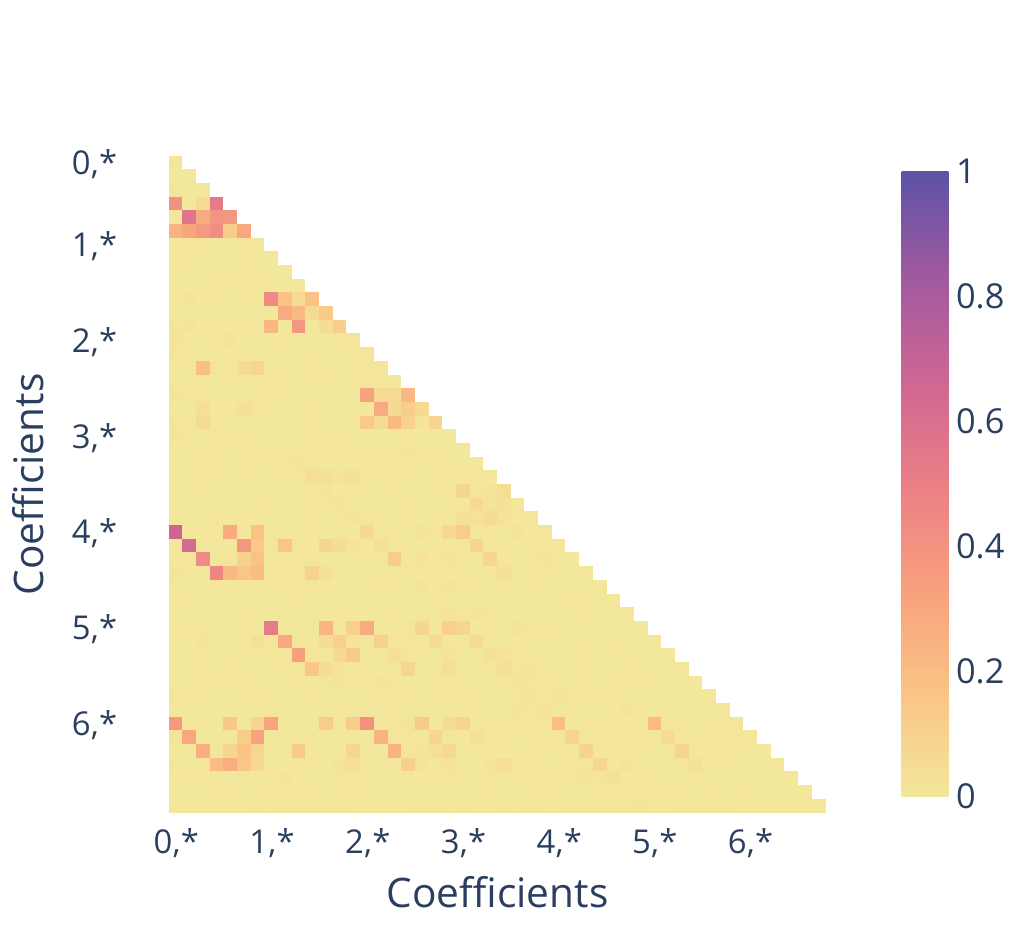}
        \caption{}
        \label{fig:fingerprints_2d}
    \end{subfigure}
    \caption{Fourier fingerprints visualizing the correlation between different Fourier coefficients (index as tick labels) of \glspl{qfm}. Circuits with the name \emph{C\_\*} are from Ref.~\cite{sim_expressibility_2019}, whilst the rest of the \ase are defined in~\autoref{app:ansatzes}. (a) Comparison between several different \ase for a 1D model. (b) Fingerprint of a 2D model for the \gls{hea} only. Here, the index reads $c_{i,\*}=\{c_{i,0}\dots c_{i,6}\}$.}
    \label{fig:fingerprints}
\end{figure}

We compute this correlation specifically for an $n=6$, $L=1$ \gls{qfm} with the 1D feature map $S = \bigotimes^n R^{(n)}_Y(x)$ and observable $\mathcal{M} = \frac{1}{n}\sum^n_i \sigma^z_i$.~\autoref{fig:fingerprints_1d_ry} shows a visualization of the computed correlations for a range of \ase.
It can be observed, that \eg the \gls{hea} results in a \gls{qfm} with significantly larger correlations between high-frequency indices whereas \emph{Circuit 15} has a minimal correlation across all coefficients.
Generally, the distribution of correlations differs significantly between the \ase, prompting us to coin the term \emph{Fourier fingerprint} to describe this unique distribution for each \as.
Note that the axes are shifted by one coefficient index as we only show the lower triangle of the correlation matrix, \ie the unique correlation without the trivial diagonal.

Furthermore, the concept of correlations between Fourier coefficients can be extended to higher input dimensions $D$ as stated in \autoref{eq:set_of_frequencies}.
Here, $\bomega$ becomes a vector of frequencies and each coefficient is indexed by $c_{(\omega_0,\omega_1)}$ for the case of $d=2$.
The correlation is then computed, similar to the 1D case, but taking into account each possible combination of frequencies, resulting in~\autoref{fig:fingerprints_2d} where we visualize the fingerprint of a single \as for a 2D model, with feature map $S = \bigotimes^n R^{(n)}_X(x_1)\bigotimes^n R^{(n)}_Y(x_2)$.
In this work, we use $500\cdot\sizeparams\cdot 2^n \cdot D$ samples to compute the correlations to high accuracy.
However, we stress that to measure the correlation to a fixed error, the number of samples does not need to grow exponentially with $n$ as shown in~\autoref{app:sampling}.

Based on these fingerprints, we introduce the \emph{\glsxtrfull{fcc}} in~\autoref{eq:fcc} as a metric that quantifies the average correlation between the coefficients.
\begin{equation}
    \fcc \coloneq \frac{1}{\vert \bOmega \vert} \sum_{\bomega, \bomega' \in \bOmega} \left\vert r_\Theta(\bomega, \bomega') \right\vert.
    \label{eq:fcc}
\end{equation}

\section{Results}
\label{sec:results}

\subsection{Learning random Fourier series}
\label{sec:results:random_fourier_series}

To investigate the predictive power of the \gls{fcc} metric for \as performance, we train a model on a truncated Fourier series $\hat f$ such that it minimizes the \gls{mse} loss
\begin{equation}
    \loss_\text{MSE}=\frac{1}{\sizex} \int_{\bx \in \mathcal{X}} \vert f(\bx, \btheta) - \hat{f}(\bx) \vert^2 d\bx,
    \label{eq:loss}
\end{equation}
where the data comes from the family of random Fourier series
\begin{equation}
    \hat{f}(\bx) = \sum_{\bomega \in \hat{\bOmega}} \hat{c}_{\bomega} \exp{(\imag \bomega \bx)}.
    \label{eq:fourier_data}
\end{equation}
Here, $\hat{c}_{\bomega}$ are uniformly drawn complex numbers within the unit circle, \ie $\sqrt{r}e^{-i2\pi p}$ with $r,p \in \mathcal{U}(0,1]$, satisfying $\vert \hat{c}_{\bomega} \vert \leq 1$.
    The frequencies of the dataset $\hat{\bOmega}$ are explicitly set to the same frequencies as those generated by the \gls{fm} of the \gls{qfm} models $\bOmega$.
    This ensures that the target function is within the learnable domain of the \glspl{qfm} assuming perfectly independent coefficients.
    Furthermore, the inputs $\bx$ are sampled according to the Nyquist-Shannon sampling theorem~\cite{shannon_communication_1949}, \ie $\sizex = \nscrit$ in the interval $[0, 2\pi)$.

We evaluate the \gls{mse} across multiple randomly seeded models and Fourier coefficients and average over the results to derive a performance index.
Specifically, we use \num{10} different model initialization and \num{10} different dataset seeds, yielding a total of \num{100} runs per \as.
We compute the expressibility using the QML-Essentials package~\cite{strobl_qml_2025}, with $500\cdot\sizeparams\cdot 2^n$ parameter samples and \num{75} histogram bins.
In~\autoref{app:computing_the_complexity_estimate} we provide details on the computation of the expressibility and a comparison to the \gls{fcc} in terms of complexity. Here we find that computing the \gls{fcc} incurs a lower overhead both on classical and quantum hardware, making it a more scalable metric for comparing \ase.

\begin{figure}[htb]
    \begin{subfigure}[c]{0.42\textwidth}
        \includegraphics[clip, trim=0.0cm 0.5cm 0.5cm 2.0cm, width=\textwidth]{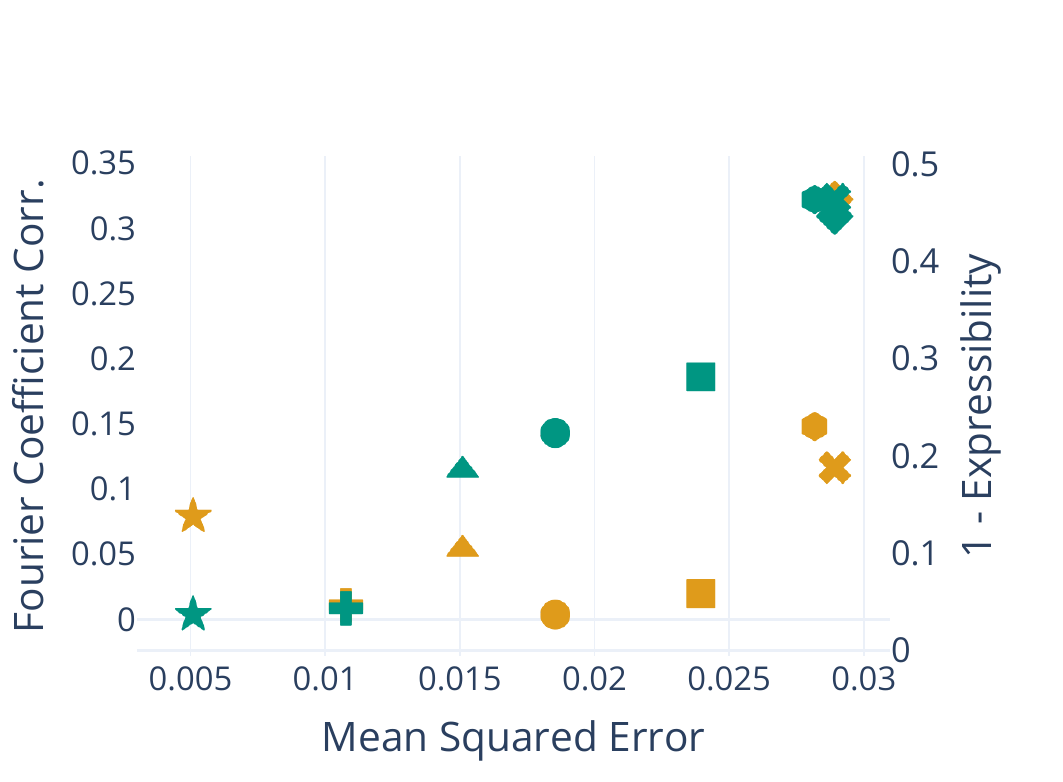}
        \caption{}
        \label{fig:mse_fcc_fs_1d_ry}
    \end{subfigure}
    \begin{subfigure}[c]{0.42\textwidth}
        \includegraphics[clip, trim=0.0cm 0.5cm 0.5cm 2.0cm, width=\textwidth]{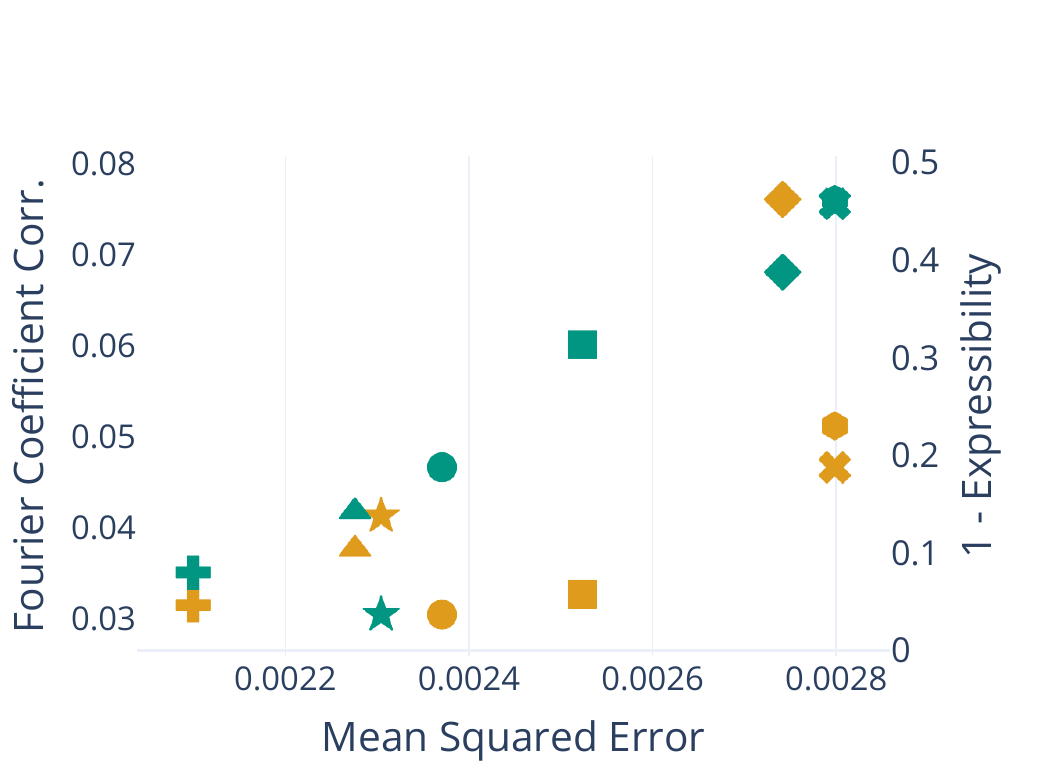}
        \caption{}
        \label{fig:mse_fcc_fs_2d}
    \end{subfigure}
    \begin{subfigure}[c]{0.14\textwidth}
        \includegraphics[clip, trim=2cm 0cm 10.7cm 2.5cm, width=\textwidth]{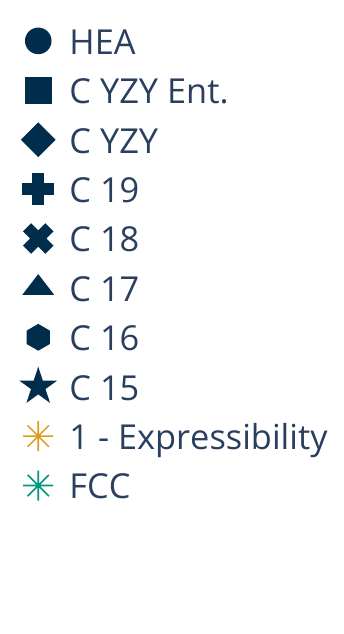}
    \end{subfigure}
    \caption{Comparison between the \textcolor{color1}{FCC} and \textcolor{color2}{expressibility complement} of models with different \ase versus the averaged \glsxtrfull{mse} when fitting Fourier series. Here, each model is trained from \num{10} random initializations and evaluated against \num{10} different (a) 1D and (b) 2D Fourier series with random coefficients.}
    \label{fig:mse_fcc_fs}
\end{figure}

The results for the 1D case are visualized in~\autoref{fig:mse_fcc_fs_1d_ry}. Here, the \gls{fcc} and expressibility are represented by different colors, whereas each symbol indicates a different \gls{qml} \as.
In line with the observations from~\autoref{fig:fingerprints_1d_ry}, \emph{Circuit 19} shows a similar FCC value to \emph{Circuit 15}, but the resulting \gls{mse} is slightly higher.
For \emph{Circuit YZY} and \emph{Circuit 18}, the \glspl{mse} are almost identical, while the \glspl{fcc} deviate slightly.
Following this, we observe that, overall, the \gls{fcc} correlates almost linearly with the \gls{mse}, with the lowest and highest values of the \gls{fcc} being \num{4.57e-3} (\emph{Circuit 15}) and \num{0.323} (\emph{Circuit 18}), respectively.
The expressibility, in contrast, does not allow for a conclusive statement and fails to correctly predict the order of the final \gls{mse} for every \as.

Employing a second Pauli encoding gate, orthogonal to the first, we replicate this training on a two-dimensional Fourier series, subsequently calculating the \gls{fcc}.
The Fourier series dataset is then naturally extended to a two-dimensional dataset, yielding the results shown in~\autoref{fig:mse_fcc_fs_2d}.

Here, one can see that \emph{Circuit 15} is now an outlier in the otherwise almost linear relationship between \gls{mse} and \gls{fcc}.
This outlier can be explained by the fact that \emph{Circuit 15} lacks the imaginary part of the coefficients when being joint by a Pauli-RX feature map, which we used to encode the second input dimension (\cf~\autoref{app:variance}).
The absence of an imaginary part depending on the rotational axis was also shown in~\cite{franz_out_2025} for this specific circuit and is included in~\autoref{app:variance}.

Overall, the results on the 1D and 2D Fourier series reveal a clear relationship between the \gls{mse} and the \gls{fcc}, indicating that this metric provides a more powerful tool for predicting model performance in these cases. Importantly, in both sets of experiments, the widely-used expressibility metric incorrectly identifies which \ase would perform the best.
We note that the standard deviations are significantly smaller than the plotted values for all measures and are mostly consistent across all \ase (\cf \autoref{app:num_errors}).

\subsection{Application to High-Energy Physics}
\label{sec:results:hep}

\Gls{hep} is a widely studied application and benchmark for \gls{qml}~\cite{schwaegerl_particle_2023,tueysuez_particle_2020,guan_quantum_2021}. In this section, we apply the same methodology from~\autoref{sec:results:random_fourier_series} to the more challenging \gls{hep} problem of 2D jet reconstruction.

Specifically, we train a \gls{qfm} to reconstruct the highest transverse momentum $p_T$ (\ie the transverse momentum of the leading jet) using the four-vector ($E$, $p_x$, $p_y$, $p_z$) of two colliding particles. 
In each of the synthetically generated collision events, two protons collide along the $z$-axis, where their constituent partons interact and produce, through a series of further interactions, decays and final hadronization, sprays of particles observed by the detector as a cluster of hadron jets.
Since the partons travel exclusively along $z$ and the LHC detector is symmetric about this axis, only the transverse momenta components in the $x$ and $y$ directions are relevant.
In the following numerical experiments, we utilize the dataset published in Ref.~\cite{haddad_parton_2025}.

\begin{figure}
    \centering
    \includegraphics[width=\linewidth]{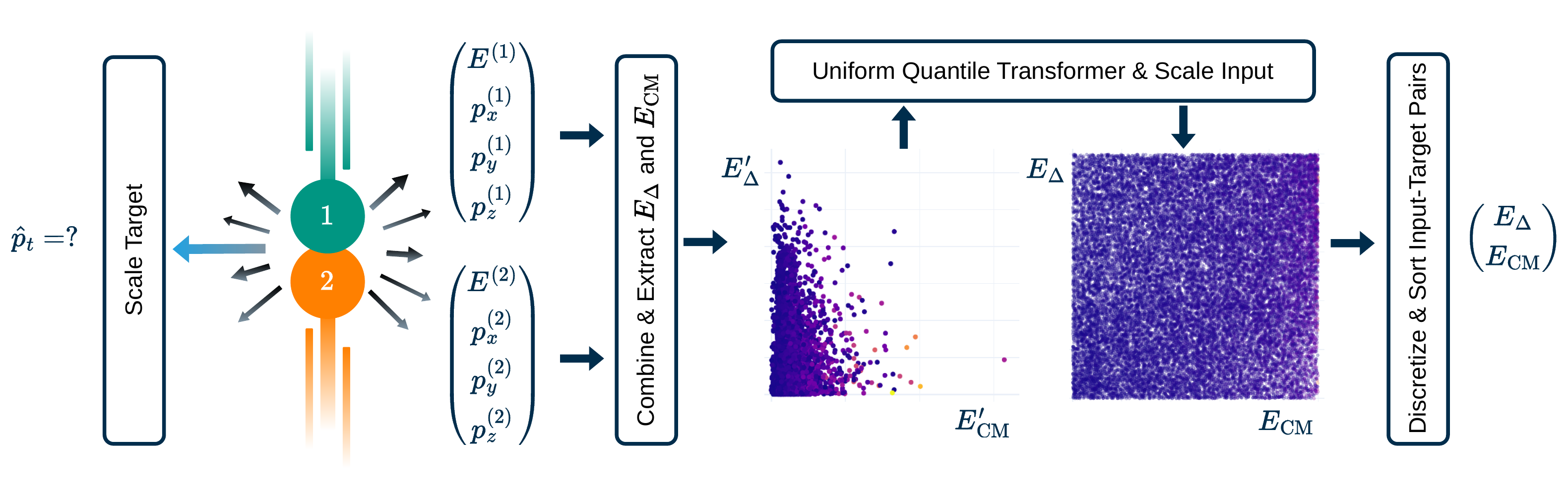}
    \caption{Problem definition and data mapping for the \gls{hep} dataset. The colliding particles $(1)$ and $(2)$ are characterized by the four vector $(E, p_x, p_y, p_z)$, which are then transformed into the center-of-mass energy $E_{\text{CM}}$ and the energy difference $E_\Delta = \left\vert E^{(1)} - E^{(2)}\right\vert$. Afterwards a  \enquote{uniform quantile transformer} is applied to the features, which are then discretized and used to predict the scaled transverse momentum $p_T$ of the leading jet.}
    \label{fig:hep_problem}
\end{figure}

In contrast to~\autoref{sec:results:random_fourier_series}, our assumption that the frequencies of the dataset match those of the \gls{qfm} can't be guaranteed in this case, making this problem potentially harder for the model to learn.
We first analyze the dataset to identify a subset of the most relevant input features for prediction. We find that the type of particle and charge have negligible influence on the leading $p_T$, which aligns with physical expectations. 
Meanwhile, we identify two input features that can be computed from the data that have a large effect on the leading $p_T$, the center-of-mass energy $E_{\text{CM}} = \sqrt{\left(E^{(1)} + E^{(2)}\right)^2 - \left(p^{(1)}_{z} + p^{(2)}_{z}\right)^2}$ and the absolute energy difference $E_\Delta = \left\vert E^{(1)} - E^{(2)}\right\vert$.
Note that due to the aforementioned symmetry, $p_x=p_y=0$. 
On these features, we apply a uniform quantile transformer to $E_{\text{CM}}$ and $E_\Delta$ resulting in a uniform distribution of the input features in the interval $[0, 2\pi)$. We then discretize this distribution to ensure equally spaced input features as conceptually visualized in~\autoref{fig:hep_problem}.

We first benchmark the complexity of this problem by applying a classical \gls{ml} model. We train a classical \gls{mlp} consisting of $2$ hidden layers and $8$ neurons per layer, resulting in $105$ trainable parameters, on the supervised learning task of predicting $p_T$ from the input features.
While being small in size, the number of parameters of this classical model still exceeds those of the \gls{qfm} tested in this work.
We train the classical model with a batch size of $256$ on a combination of the \gls{kl} divergence (KLDiv) and the Huber loss as $\loss = \text{KLDiv} + 0.001 \cdot \text{HuberLoss}$ using the Adam optimizer~\cite{kingma_adam_2017} with a learning rate of $0.05$, adjusted dynamically via a learning rate scheduler.
We justify the choice of this loss function as we aim to learn the momenta distribution of the dataset, due to the probabilistic nature of the task and no one-to-one input to output mapping, which \text{KLDiv} naturally models.
Input features are scaled using \enquote{MinMax} scaling, while the leading $p_T$ is left unscaled.
KL-divergence is used exclusively for validation and testing, while we add a small Huber loss during training, as it empirically improved convergence stability.

Training is performed on \num{200000} events with a ratio of $80\%$ for training, $10\%$ for validation and $10\%$ for testing.
On average, across \num{10} random models, the classical \gls{mlp} obtained a \gls{kl} divergence score of $0.163 \pm 0.049$.
However, selecting the \num{10} best-performing models from a pool of \num{30} improved the result significantly, yielding a \gls{kl} divergence of $0.035 \pm 0.005$.
Despite trying multiple engineering tactics - including varying model architecture, model size, and learning strategies - performance did not improve.
Consequently, we selected the simplest solution that still yielded competitive results.

Following an evaluation of the classical model, we shift our attention to the \gls{qfm}.
As in previous experiments described in~\autoref{sec:results:random_fourier_series}, we use \num{10} different model initializations and \num{10} different dataset initializations.
In contrast to the preliminary experiments on \gls{mlp}, we reduce the dataset size to \num{40000} with a ratio of $80\%$ for training, $10\%$ for validation and $10\%$ for testing, both for any subsequent experiments on the \gls{mlp} and \gls{qfm}.
We switch the previous \enquote{MinMax} scaler of the classical \gls{mlp} to a \enquote{uniform quantile transformer} as in the \gls{qfm}.
Similarly, the target values using \enquote{MinMax} scaling and compared with the predicted values combining the \gls{mse} and KL-divergence loss~\cite{kullback_information_1951} $\loss = \text{MSE} + 0.001 \cdot \text{KLDiv}$ yields the best results in conjunction with an Adam optimizer~\cite{kingma_adam_2017} with a learning rate of $0.005$.

\begin{figure}[htb]
    \begin{subfigure}[c]{0.42\textwidth}
        \includegraphics[clip, trim=0.0cm 0.5cm 1.0cm 2.0cm, width=\textwidth]{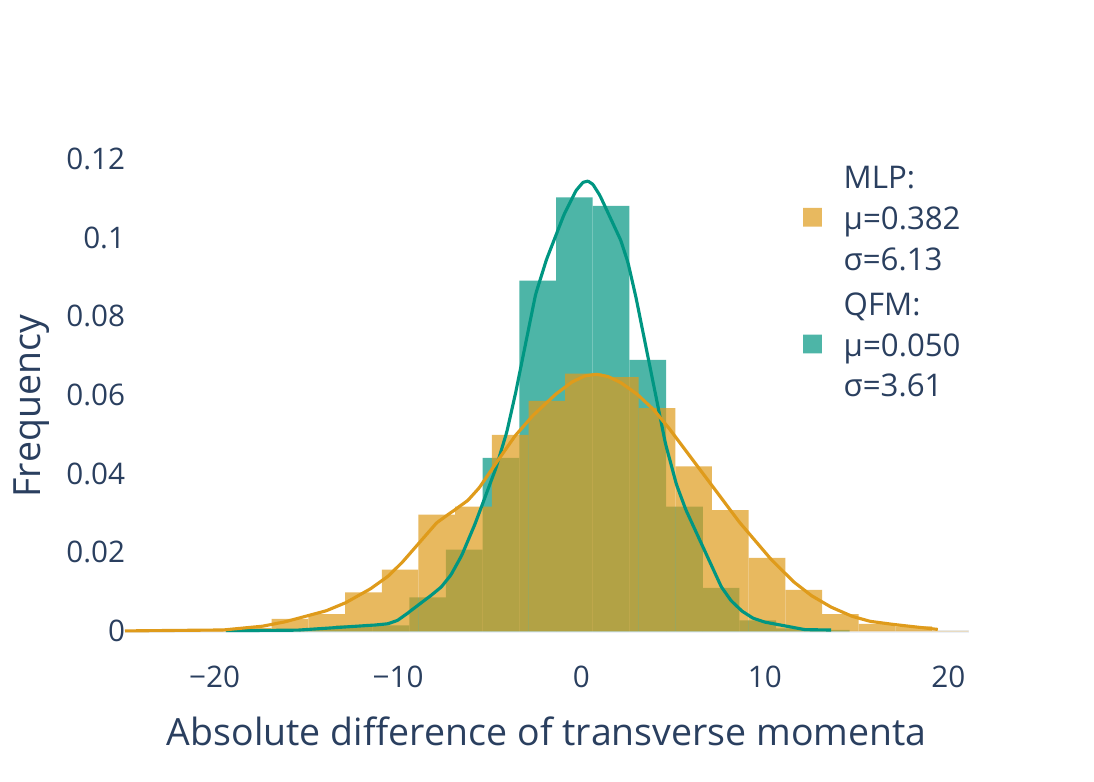}
        \caption{}
        \label{fig:distributions_hep_2d}
    \end{subfigure}
    \begin{subfigure}[c]{0.42\textwidth}
        \includegraphics[clip, trim=0.0cm 0.5cm 0.5cm 2.0cm, width=\textwidth]{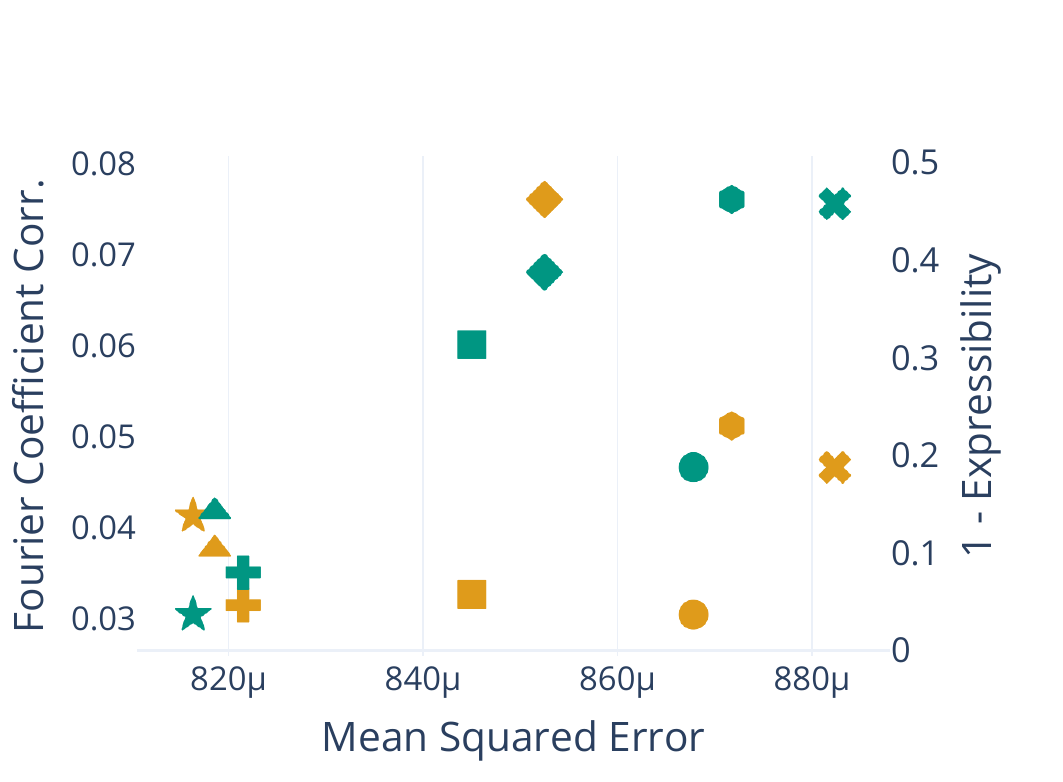}
        \caption{}
        \label{fig:mse_fcc_hep_2d}
    \end{subfigure}
    \begin{subfigure}[c]{0.14\textwidth}
        \includegraphics[clip, trim=2cm 0cm 10.7cm 2.5cm, width=\textwidth]{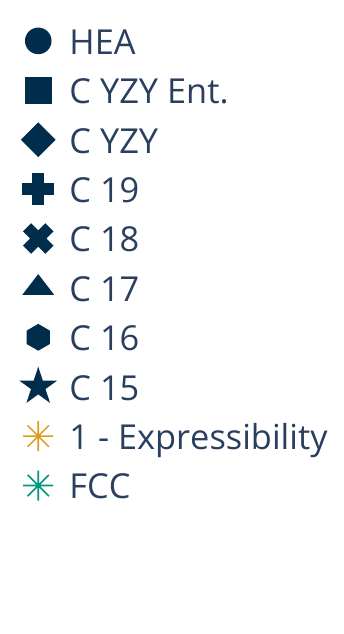}
    \end{subfigure}
    \caption{Training performance (validation score) of (a): the \gls{mlp} and \gls{qfm} visualized as distributions of the differences between the target $p_T$ and the ground truth and (b): Training performance (validation score) of the model on the \gls{hep} dataset. Averaged over \num{10} independent dataset seeds and \num{10} distinct parameter initializations.}
    \label{fig:hep_results}
\end{figure}

A comparison between the results obtained by the \gls{mlp} and \gls{qfm} is shown in~\autoref{fig:distributions_hep_2d}. Specifically, we plot the distribution of absolute differences between the predicted and true $p_T$ values in each case.
Here, we observe that both models achieve a low mean deviation of $0.382$ and $0.050$ for the \gls{mlp} and \gls{qfm}, respectively.
However, the standard deviation of the absolute differences is lower for the \gls{qfm} with a standard deviation of $3.61$ compared to $6.13$ for the \gls{mlp}.
Overall, our results validate that the problem is non-trivial to solve with a standard \gls{mlp}, and that our trained \gls{qfm} produces a competitive benchmark.

In~\autoref{fig:mse_fcc_hep_2d} we show the comparison of the predicted performance of the \gls{fcc} and expressibility for the \gls{hep} dataset.
When observing the \gls{fcc}, \emph{Circuit 17} and the \enquote{Hardware Efficient Ansatz} can be identified as outliers.
The \as with the lowest and highest \gls{fcc}, \emph{Circuit 15} and \emph{Circuit 18}, correlate to the lowest and highest \gls{mse}, respectively.
Although the order of \glspl{mse} is not perfectly predicted by the \glspl{fcc} here, we observe a clear overall trend linking correlation and performance.
Compared to expressibility, both metrics successfully predict the same clusters of \ase while the \gls{fcc} is computationally less expensive as discussed in~\autoref{app:computing_the_complexity_estimate}.
In addition to the results considering the \gls{mse}, we provide the results for the \gls{kl} divergence and Huber loss in~\autoref{app:high-energy-physics}.

\section{Conclusion \& discussion}

In this work, we have demonstrated the existence of correlations between the Fourier coefficients of \glspl{qfm} and its impact on performance for several popular \ase.
In~\autoref{sec:method}, we introduced the theoretical motivation as to why these correlations exist and how they can be numerically calculated by repeatedly sampling random instances of the model.
In~\autoref{sec:results:random_fourier_series}, we showed how, for the task of learning random Fourier series, models with lower average \gls{fcc} consistently perform better for both 1D and 2D inputs.
Finally, in~\autoref{sec:results:hep} we evidenced how this relationship extends to problems of practical interest such as the more challenging setting of 2D jet reconstruction in high-energy physics.

Despite our numerical results, we note that correlations between the Fourier coefficients of efficiently-trainable \glspl{qfm} are only asymptotically guaranteed for feature maps that produce exponentially many \emph{unique} frequencies.
This includes the Goulomb encoding~\cite{peters_generalization_2022} as well as those in Refs.~\cite{shin_exponential_2023, kyriienko_protocols_2024}.
Whilst the feature maps used in this work produce \glspl{qfm} with exponentially many basis functions, in contrast, there are only a linear number of unique modes due to degeneracy in the eigenspectrum.
This implies that it may be possible to design an \as that, in combination with a degenerate feature map, could have fully independent Fourier coefficients.
This could be achieved by ensuring each parameter only appears in the coefficients of terms with the same frequency. Nevertheless, our results highlight that this phenomenon does not occur in popular \ase.
Understanding what such an \as would look like, and if it is efficiently implementable, remains an interesting open question.

Following this, the aspect of trainability should also be considered an important area of future research.
\citeauthor{ragone_lie_2024}~\cite{ragone_lie_2024} showed that \ase with high expressibility are very unlikely to be trainable, raising the interesting question of whether a similar pattern translates to the \gls{fcc}.
The numerical results presented in this paper do not strictly imply that a high expressibility correlates with a low \gls{fcc} (\cf \autoref{fig:mse_fcc_fs}), suggesting a non-trivial relationship between these metrics.
Therefore, it cannot be ruled out that a lower \gls{fcc}, \ie \ase with highly-independent coefficients, are generally not trainable.

Finally, in this work, we studied only one instance of a real-world dataset. As a future direction of research, it remains to be seen whether the same trends hold for other problems, whereby \ase with the lowest average \gls{fcc} perform the best.
In particular, this heuristic has the underlying assumption that the frequencies of the dataset are themselves independent.
If this is not true, one may find that the best \gls{qml} model is one which purposefully produces Fourier coefficients with large correlations specifically tied to the data.
In this way, the \gls{fcc} may be better thought of as an inductive bias to encode into \glspl{qfm}, instead of as a metric to always minimize.

\section*{Acknowledgements}

We thank Gabriel Mejía Ruiz for helpful discussions on mathematical formulations and Maja Franz for valuable discussions concerning \glspl{qfm} in a broader view.

We want to express our gratitude to Ryan Sweke and Christa Zoufal for their useful discussions and insights.

MS, EK and AS acknowledge support by the state of Baden-W\"urttemberg through bwHPC.

This work was supported by the Hartree National Centre for Digital Innovation, a collaboration between the Science and Technology Facilities Council and IBM.

\printbibliography

\clearpage
\glsresetall
\appendix

\section{\Ase}
\label{app:ansatzes}

\begin{figure}[htb]
    \centering
    \begin{subfigure}[c]{0.26\textwidth}
        \centering
        \begin{tikzpicture}
            \node[scale=0.72] {
                \scriptsize\begin{quantikz}[row sep=0.2cm, column sep=0.7mm]
                    \qw & \gate{RY(\theta_{0})} & \gate{RZ(\theta_{1})} & \gate{RY(\theta_{2})} & \qw \\
                    \qw & \gate{RY(\theta_{3})} & \gate{RZ(\theta_{4})} & \gate{RY(\theta_{5})} & \qw \\
                    \qw & \gate{RY(\theta_{6})} & \gate{RZ(\theta_{7})} & \gate{RY(\theta_{8})} & \qw \\
                    \qw & \gate{RY(\theta_{9})} & \gate{RZ(\theta_{10})} & \gate{RY(\theta_{11})} & \qw
                \end{quantikz}
            };
        \end{tikzpicture}
        \caption{Circuit YZY.}
        \label{fig:circuit_yzy}
    \end{subfigure}
    \begin{subfigure}[c]{0.36\textwidth}
        \centering
        \begin{tikzpicture}
            \node[scale=0.72] {
                \scriptsize\begin{quantikz}[row sep=0.2cm, column sep=0.7mm]
                    \qw & \gate{RY(\theta_{0})} & \gate{RZ(\theta_{1})} & \gate{RY(\theta_{2})} & \ctrl{1} & \ctrl{2} & \ctrl{3} &  &  &  & \qw \\
                    \qw & \gate{RY(\theta_{3})} & \gate{RZ(\theta_{4})} & \gate{RY(\theta_{5})} & \targ{} &  &  & \ctrl{1} & \ctrl{2} &  & \qw \\
                    \qw & \gate{RY(\theta_{6})} & \gate{RZ(\theta_{7})} & \gate{RY(\theta_{8})} &  & \targ{} &  & \targ{} &  & \ctrl{1} & \qw \\
                    \qw & \gate{RY(\theta_{9})} & \gate{RZ(\theta_{10})} & \gate{RY(\theta_{11})} &  &  & \targ{} &  & \targ{} & \targ{} & \qw
                \end{quantikz}
            };
        \end{tikzpicture}
        \caption{Circuit YZY Entangling.}
        \label{fig:circuit_yzy_entangling}
    \end{subfigure}
    \begin{subfigure}[c]{0.32\textwidth}
        \centering
        \begin{tikzpicture}
            \node[scale=0.72] {
                \scriptsize\begin{quantikz}[row sep=0.2cm, column sep=0.7mm]
                    \qw & \gate{RY(\theta_{0})} & \gate{RZ(\theta_{1})} & \gate{RY(\theta_{2})} & \ctrl{1} &  & \targ{} & \qw \\
                    \qw & \gate{RY(\theta_{3})} & \gate{RZ(\theta_{4})} & \gate{RY(\theta_{5})} & \targ{} & \ctrl{1} &  & \qw \\
                    \qw & \gate{RY(\theta_{6})} & \gate{RZ(\theta_{7})} & \gate{RY(\theta_{8})} & \ctrl{1} & \targ{} &  & \qw \\
                    \qw & \gate{RY(\theta_{9})} & \gate{RZ(\theta_{10})} & \gate{RY(\theta_{11})} & \targ{} &  & \ctrl{-3} & \qw
                \end{quantikz}
            };
        \end{tikzpicture}
        \caption{Hardware-efficient \as.}
        \label{fig:hardware_efficient_ansatz}
    \end{subfigure}\\
    \begin{subfigure}[c]{0.39\textwidth}
        \centering
        \begin{tikzpicture}
            \node[scale=0.72] {
                \scriptsize\begin{quantikz}[row sep=0.2cm, column sep=0.7mm]
                    \qw & \gate{RY(\theta_{0})} & \targ{} &  &  & \ctrl{1} & \gate{RY(\theta_{4})} & \ctrl{3} & \targ{} &  & \qw \\
                    \qw & \gate{RY(\theta_{1})} &  &  & \ctrl{1} & \targ{} & \gate{RY(\theta_{5})} &  & \ctrl{-1} & \targ{} & \qw \\
                    \qw & \gate{RY(\theta_{2})} &  & \ctrl{1} & \targ{} & \gate{RY(\theta_{6})} & \targ{} &  &  & \ctrl{-1} & \qw \\
                    \qw & \gate{RY(\theta_{3})} & \ctrl{-3} & \targ{} & \gate{RY(\theta_{7})} &  & \ctrl{-1} & \targ{} &  &  & \qw
                \end{quantikz}
            };
        \end{tikzpicture}
        \caption{Circuit 15.}
        \label{fig:circuit_15}
    \end{subfigure}
    \begin{subfigure}[c]{0.29\textwidth}
        \centering
        \begin{tikzpicture}
            \node[scale=0.72] {
                \scriptsize\begin{quantikz}[row sep=0.2cm, column sep=0.7mm]
                    \qw & \gate{RX(\theta_{0})} & \gate{RZ(\theta_{1})} & \gate{RZ(\theta_{8})} &  & \qw \\
                    \qw & \gate{RX(\theta_{2})} & \gate{RZ(\theta_{3})} & \ctrl{-1} & \gate{RZ(\theta_{10})} & \qw \\
                    \qw & \gate{RX(\theta_{4})} & \gate{RZ(\theta_{5})} & \gate{RZ(\theta_{9})} & \ctrl{-1} & \qw \\
                    \qw & \gate{RX(\theta_{6})} & \gate{RZ(\theta_{7})} & \ctrl{-1} &  & \qw
                \end{quantikz}
            };
        \end{tikzpicture}
        \caption{Circuit 16.}
        \label{fig:circuit_16}
    \end{subfigure}
    \begin{subfigure}[c]{0.30\textwidth}
        \centering
        \begin{tikzpicture}
            \node[scale=0.72] {
                \scriptsize\begin{quantikz}[row sep=0.2cm, column sep=0.7mm]
                    \qw & \gate{RX(\theta_{0})} & \gate{RZ(\theta_{1})} & \gate{RX(\theta_{8})} &  & \qw \\
                    \qw & \gate{RX(\theta_{2})} & \gate{RZ(\theta_{3})} & \ctrl{-1} & \gate{RX(\theta_{10})} & \qw \\
                    \qw & \gate{RX(\theta_{4})} & \gate{RZ(\theta_{5})} & \gate{RX(\theta_{9})} & \ctrl{-1} & \qw \\
                    \qw & \gate{RX(\theta_{6})} & \gate{RZ(\theta_{7})} & \ctrl{-1} &  & \qw
                \end{quantikz}
            };
        \end{tikzpicture}
        \caption{Circuit 17.}
        \label{fig:circuit_17}
    \end{subfigure}\\
    \begin{subfigure}[c]{0.46\textwidth}
        \centering
        \begin{tikzpicture}
            \node[scale=0.72] {
                \scriptsize\begin{quantikz}[row sep=0.2cm, column sep=0.7mm]
                    \qw & \gate{RX(\theta_{0})} & \gate{RZ(\theta_{1})} & \gate{RZ(\theta_{8})} &  &  & \ctrl{1} & \qw \\
                    \qw & \gate{RX(\theta_{2})} & \gate{RZ(\theta_{3})} &  &  & \ctrl{1} & \gate{RZ(\theta_{11})} & \qw \\
                    \qw & \gate{RX(\theta_{4})} & \gate{RZ(\theta_{5})} &  & \ctrl{1} & \gate{RZ(\theta_{10})} &  & \qw \\
                    \qw & \gate{RX(\theta_{6})} & \gate{RZ(\theta_{7})} & \ctrl{-3} & \gate{RZ(\theta_{9})} &  &  & \qw
                \end{quantikz}
            };
        \end{tikzpicture}
        \caption{Circuit 18.}
        \label{fig:circuit_18}
    \end{subfigure}
    \begin{subfigure}[c]{0.46\textwidth}
        \centering
        \begin{tikzpicture}
            \node[scale=0.72] {
                \scriptsize\begin{quantikz}[row sep=0.2cm, column sep=0.7mm]
                    \qw & \gate{RX(\theta_{0})} & \gate{RZ(\theta_{1})} & \gate{RX(\theta_{8})} &  &  & \ctrl{1} & \qw \\
                    \qw & \gate{RX(\theta_{2})} & \gate{RZ(\theta_{3})} &  &  & \ctrl{1} & \gate{RX(\theta_{11})} & \qw \\
                    \qw & \gate{RX(\theta_{4})} & \gate{RZ(\theta_{5})} &  & \ctrl{1} & \gate{RX(\theta_{10})} &  & \qw \\
                    \qw & \gate{RX(\theta_{6})} & \gate{RZ(\theta_{7})} & \ctrl{-3} & \gate{RX(\theta_{9})} &  &  & \qw
                \end{quantikz}
            };
        \end{tikzpicture}
        \caption{Circuit 19.}
        \label{fig:circuit_19}
    \end{subfigure}
    \caption{The \ase used in this work, exemplarily for $n=4$ qubits. Note the circular structure of the \gls{hea}.}
    \label{fig:ansatzes}
\end{figure}
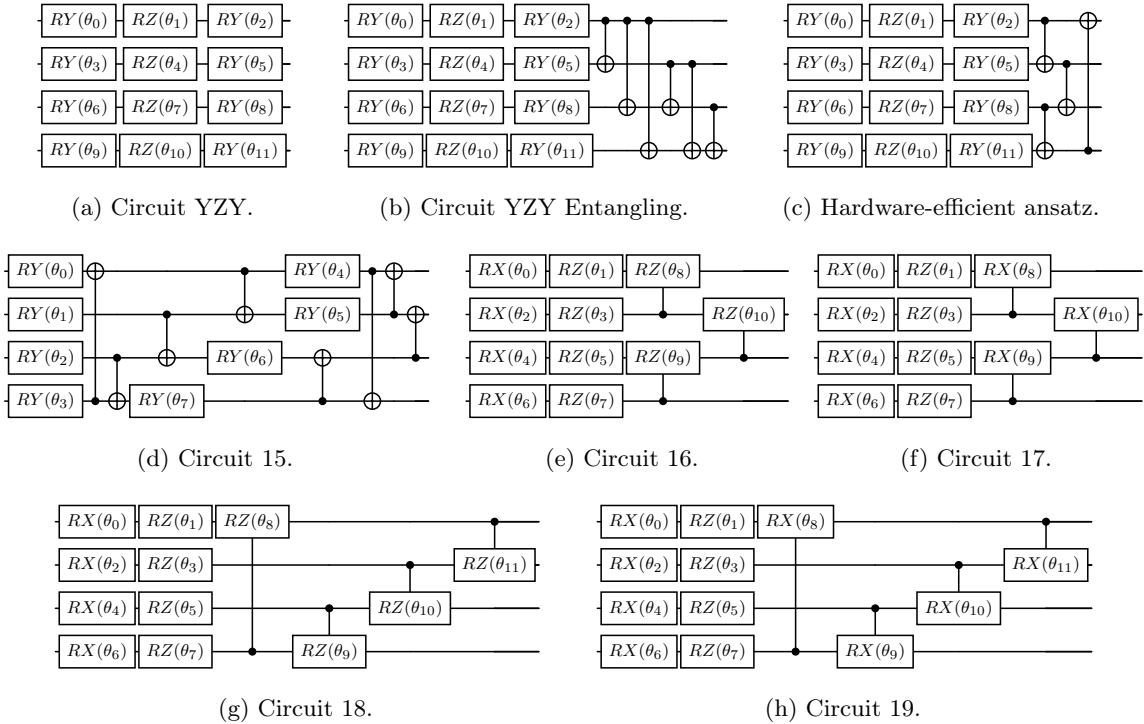

~\autoref{fig:ansatzes} shows the \ase used in this work. Specifically, we depict only a single variational layer and do not include the feature map.

\section{Effect of rotational axis on FCC}
\label{app:rotational_angle}

\begin{figure}[htb]
    \begin{subfigure}[c]{0.46\textwidth}
        \includegraphics[clip, trim=0.2cm 0.0cm 0.0cm 2.0cm, width=\textwidth]{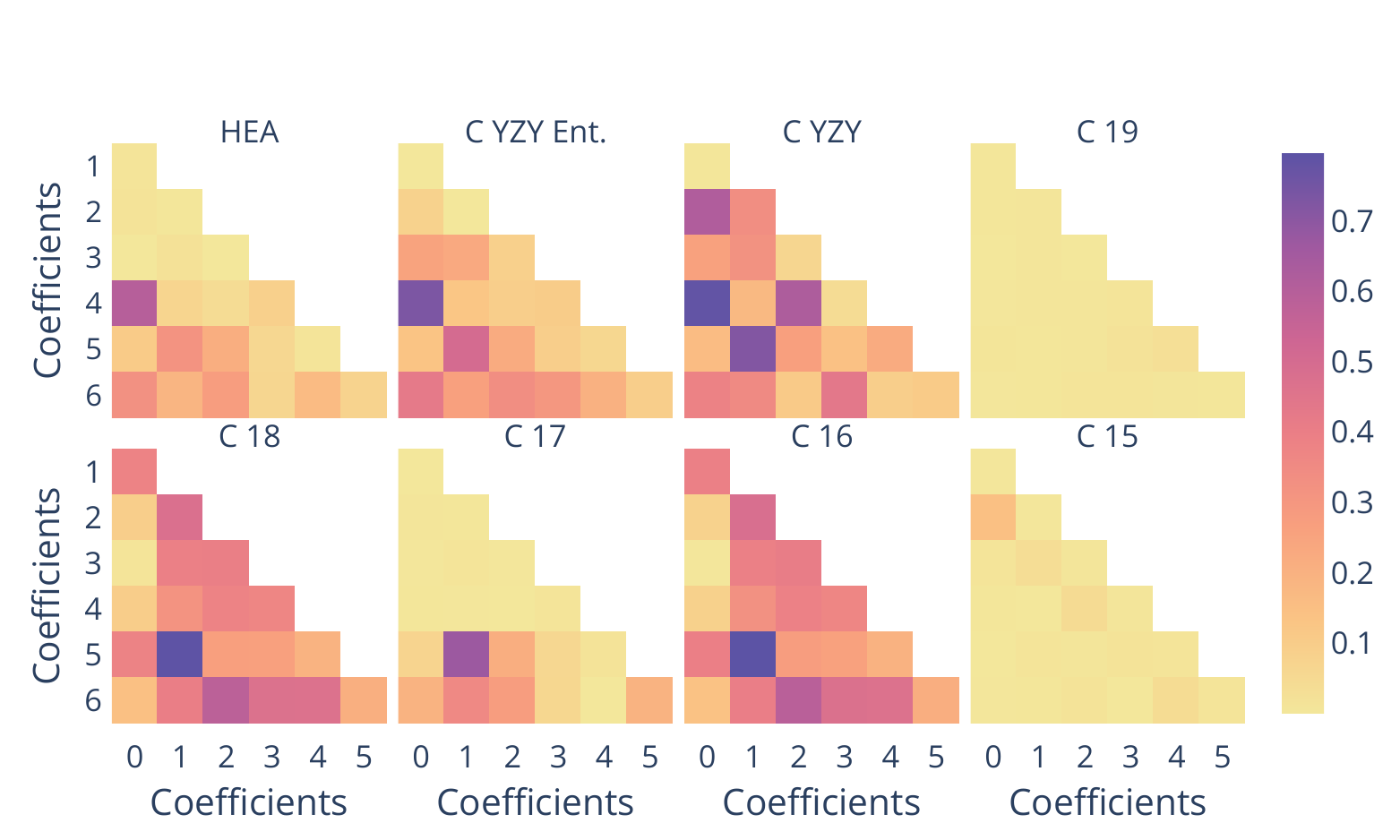}
        \caption{}
        \label{fig:fingerprints_1d_rx}
    \end{subfigure}
    \begin{subfigure}[c]{0.38\textwidth}
        \includegraphics[clip, trim=0.0cm 0.5cm 0.5cm 2.0cm, width=\textwidth]{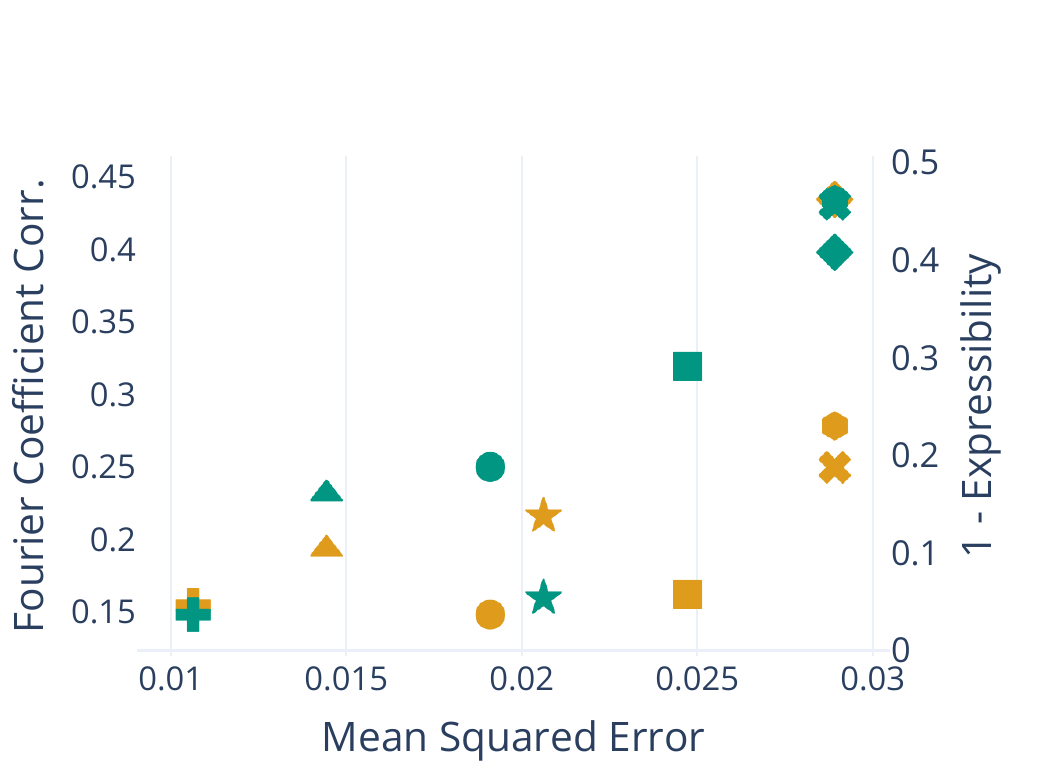}
        \caption{}
        \label{fig:mse_fcc_fs_1d_rx}
    \end{subfigure}
    \begin{subfigure}[c]{0.14\textwidth}
        \includegraphics[clip, trim=2cm 0cm 10.7cm 2.5cm, width=\textwidth]{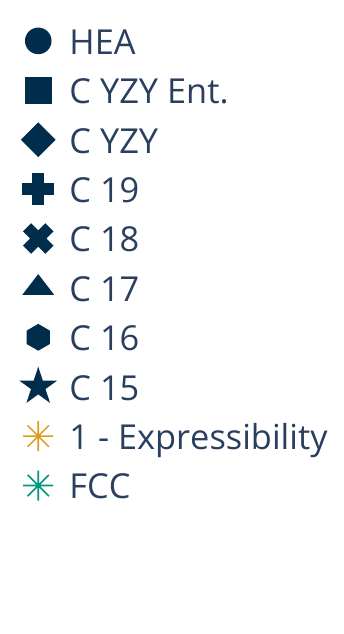}
    \end{subfigure}
    \caption{Fourier fingerprints (a) and comparison between the \textcolor{color1}{FCC} and \textcolor{color2}{expressibility complement} over the \glsxtrfull{mse} for different \ase depicted by different shapes. Training on 1D Fourier series with \num{10} independent dataset seeds and \num{10} distinct parameter initializations. Features are encoded using RX gates.}
    \label{fig:mse_fcc_fs_uw_rx}
\end{figure}

In~\autoref{fig:mse_fcc_fs_1d_ry} we utilize RY gates in the feature map of the models as we observed an absence of imaginary parts in \emph{Circuit 15} for RX encoding gates.
For the sake of completeness, we provide the Fourier fingerprints and comparison between the \gls{fcc} and expressibility in~\autoref{fig:fingerprints_1d_rx} and~\autoref{fig:mse_fcc_fs_1d_rx} respectively when the model is trained using RX encoding gates.

\section{Impact of the variance of coefficients in the FCC}
\label{app:variance}

\begin{figure}[htb]
    \begin{minipage}[c]{0.33\textwidth}
        \includegraphics[width=\textwidth]{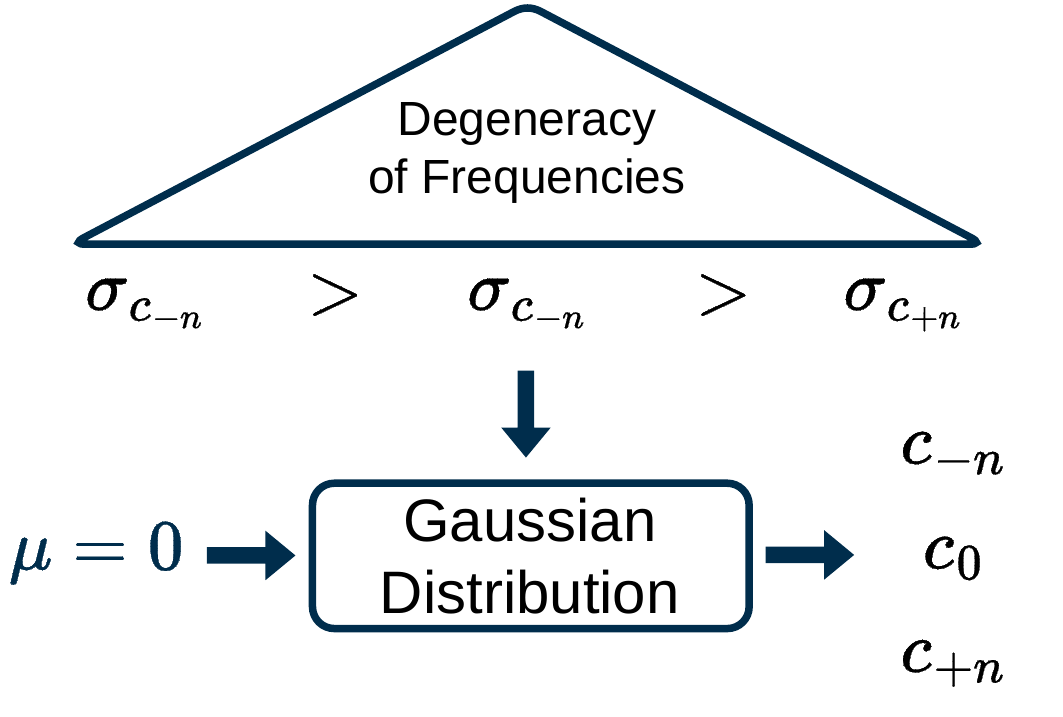}
    \end{minipage}
    \begin{minipage}[c]{0.33\textwidth}
        \includegraphics[width=\textwidth]{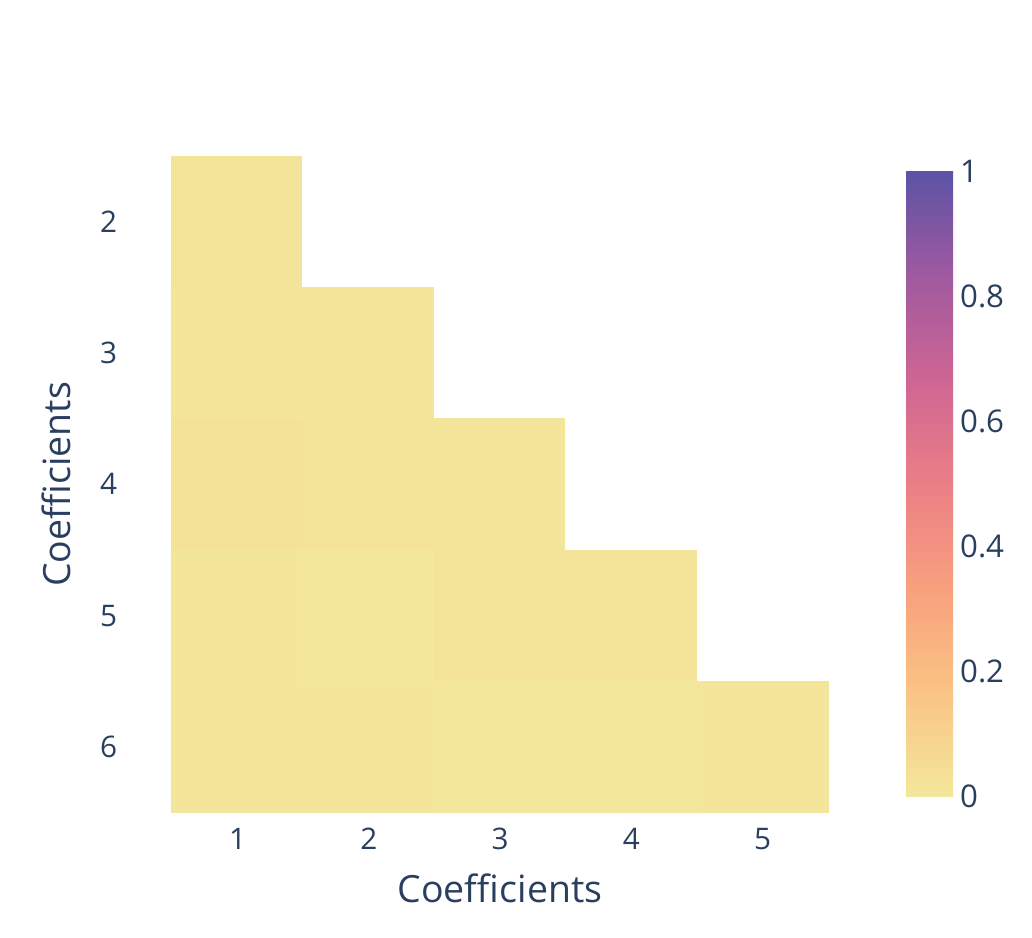}
    \end{minipage}\hfill
    \begin{minipage}[c]{0.3\textwidth}
        \caption{Correlation matrix of randomly generated coefficients samples by a Gaussian distribution with zero mean and the variance set to the degeneracy of the eigenvalues of the \gls{fm}. Visualization without redundancies and values are expected to be all-zero.}
        \label{fig:coefficient_correlation_surrogate}
    \end{minipage}
\end{figure}

The variance of the Fourier coefficients of \glspl{qfm} is linked to the redundancy of the frequency, as observed in~\cite{mhiri_constrained_2024}.
For the \glspl{fm} studied in this work, this means that the variance of the coefficients decays exponentially towards higher frequencies.
In this section, we consider whether this can lead to spurious correlations between coefficients which are independent but have the same redundancy.
This is particularly important, since the variance of the coefficients enters in the denominator of our definition of correlation in equation~\autoref{eq:pearson_correlation}.
To numerically validate this, we randomly sample a set of coefficients from a Gaussian distribution with zero mean and the variance set to the degeneracy of the eigenvalues of the \gls{fm}.
Applying the Pearson method $r$ we obtain the correlation between the coefficients samples, which are shown in~\autoref{fig:coefficient_correlation_surrogate} and can be seen to be all-zero, as expected.

\begin{figure}[htb]
    \begin{subfigure}[c]{0.38\textwidth}
        \includegraphics[clip, trim=0cm 0.5cm 1cm 2cm, width=\textwidth]{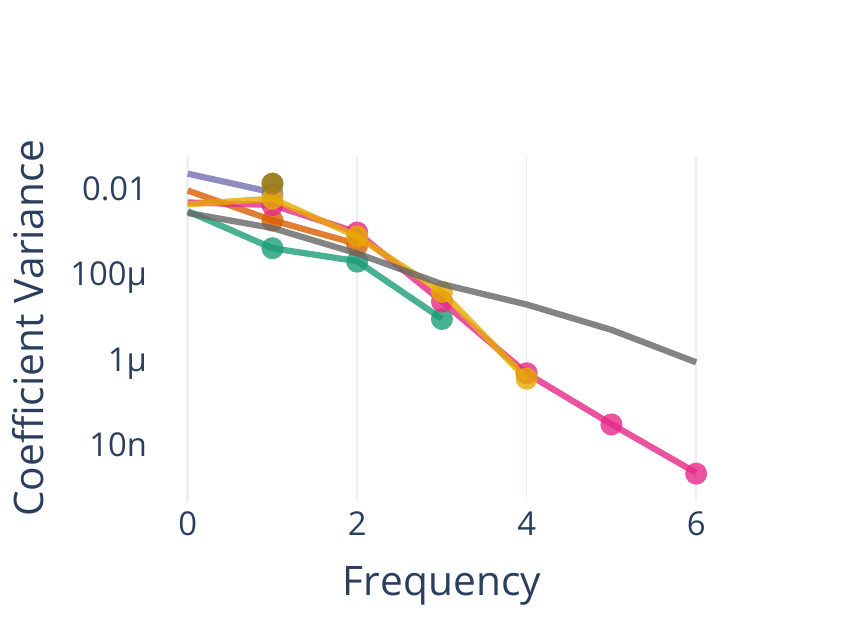}
        \caption{}
    \end{subfigure}
    \begin{subfigure}[c]{0.38\textwidth}
        \includegraphics[clip, trim=0cm 0.5cm 1cm 2cm, width=\textwidth]{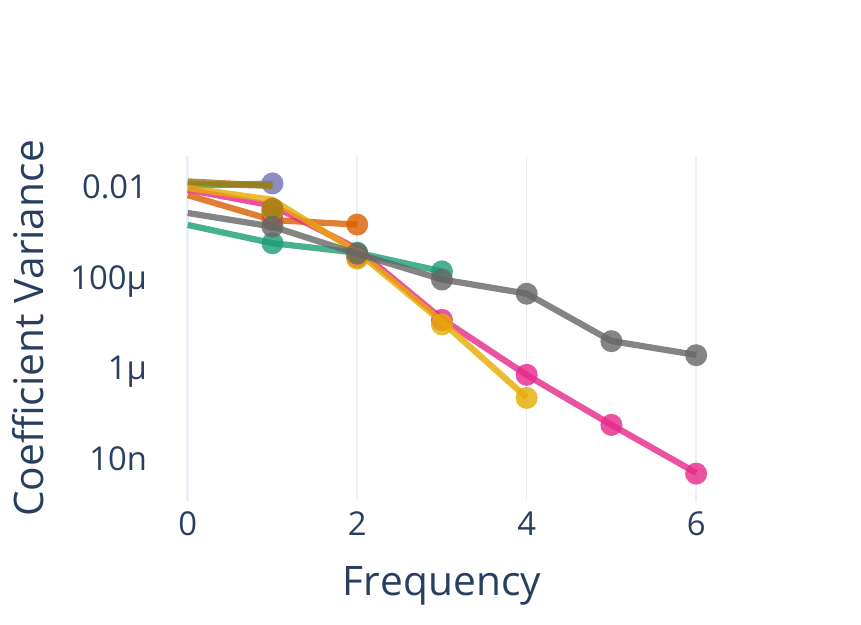}
        \caption{}
    \end{subfigure}
    \begin{subfigure}[c]{0.14\textwidth}
        \includegraphics[clip, trim=2cm 0cm 10.7cm 2.5cm, width=\textwidth]{figures/1dfs/rx/1dfs_rx_mse_uw_sce_q6_legend.pdf}
    \end{subfigure}
    \caption{Variances of the coefficient values $\var(\Re\{c_{\bomega}\})$ (lines) and $\var(\Im\{c_{\bomega}\})$ (markers) averaged over \num{10} different seeds for different frequencies $\bomega$ and different \ase with (a) RX gates and (b) RY gates. Variances are cut off at $10^{-10}$.}
    \label{fig:coefficient_variance}
\end{figure}

It is important that the \gls{fcc} is independent of the variance of the coefficients, since, in practice, this varies in magnitude at different frequencies for different circuits.
We can observe this effect in~\autoref{fig:coefficient_variance}, where we show the real- and imaginary part of the variance decaying exponentially towards higher frequencies.
The correlations are computed using $500\cdot\sizeparams\cdot 2^n \cdot D$ parameter samples with $10$ different initial seeds on $n=6$ qubit \ase.
This finding is in line with a main statement from Ref.~\cite{mhiri_constrained_2024}.
Here, specifically, we find that only \emph{Circuit 15} and \emph{Circuit 19} are exhibiting $\text{Var}(\vert c_{\bomega} \vert) > 10^{-10}$ for all frequencies evaluated.
Furthermore, we can see that \emph{Circuit 15} does not exhibit an imaginary part when utilizing RX gates for the \gls{fm}.

Notably, we observe that the mean value of the coefficients remains zero, \ie that the absolute value of the coefficients is, independent of the frequency, zero-centered.

\section{Weighting the FCC}
\label{app:weighting_the_fcc}

\begin{figure}[htb]
    \begin{subfigure}[c]{0.46\textwidth}
        \includegraphics[clip, trim=0.2cm 0.5cm 0.0cm 2.0cm, width=\textwidth]{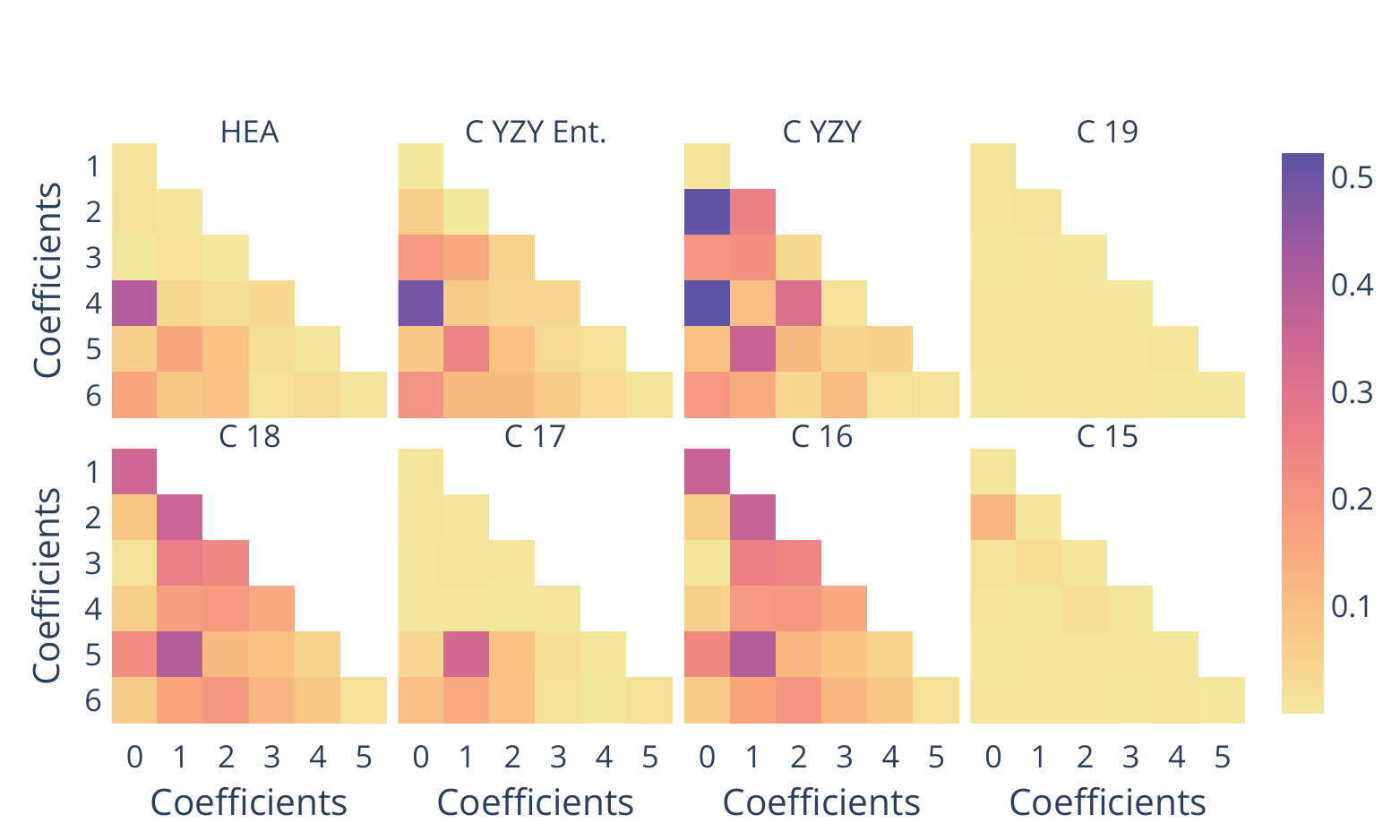}
        \caption{}
        \label{fig:fingerprints_1d_rx_w}
    \end{subfigure}
    \begin{subfigure}[c]{0.38\textwidth}
        \includegraphics[clip, trim=0.0cm 0.5cm 0.5cm 2.0cm, width=\textwidth]{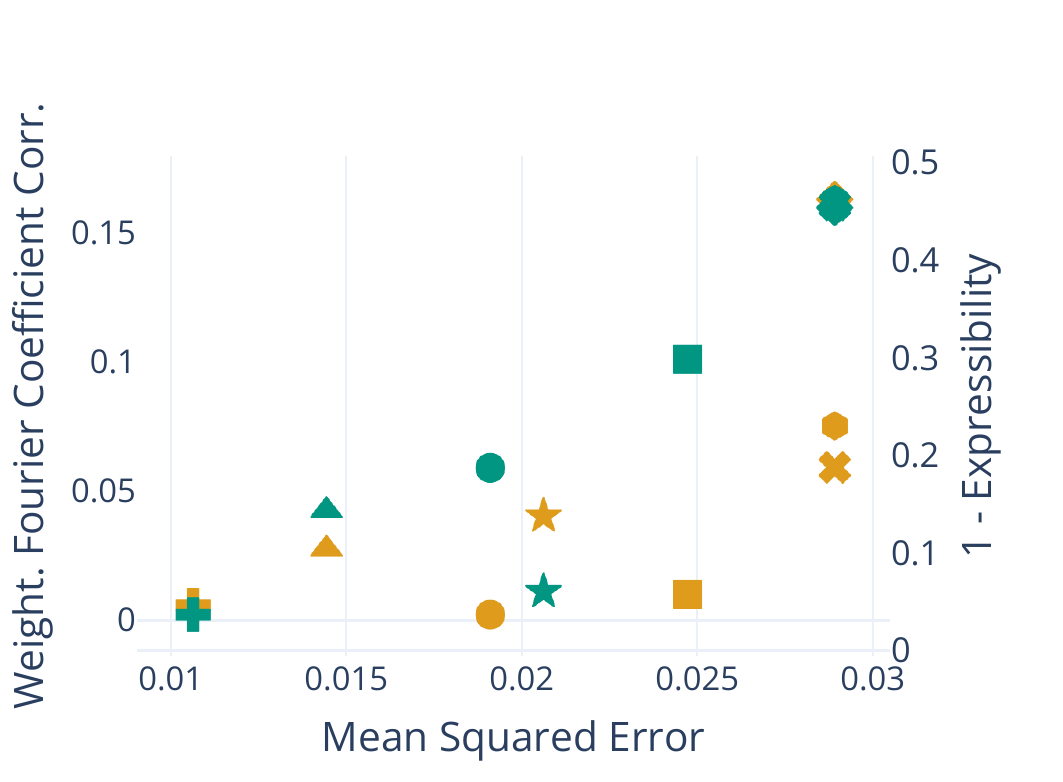}
        \caption{}
        \label{fig:mse_fcc_fs_1d_rx_w}
    \end{subfigure}
    \begin{subfigure}[c]{0.14\textwidth}
        \includegraphics[clip, trim=2cm 0cm 10.7cm 2.5cm, width=\textwidth]{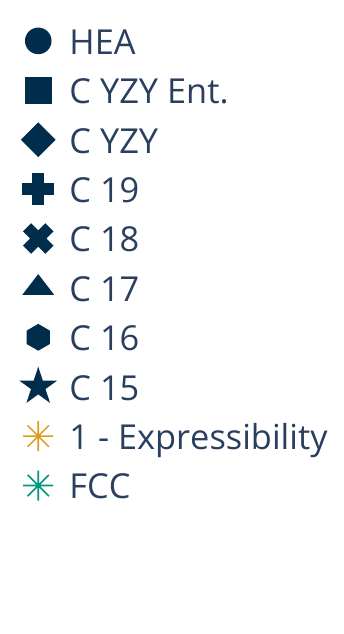}
    \end{subfigure}
    \caption{Fourier fingerprints (a) and comparison between the \textcolor{color1}{FCC} and \textcolor{color2}{expressibility complement} over the \glsxtrfull{mse} for different \ase depicted by different shapes. Training on 1D Fourier series with \num{10} independent dataset seeds and \num{10} distinct parameter initializations. Features encoded using RX gates and linear weighting strategy on the \gls{fcc}.}
    \label{fig:mse_fcc_fs_w_rx}
\end{figure}

Considering the findings of~\autoref{app:variance}, the \gls{fcc} as introduced in~\autoref{eq:fcc} does not take into account the degeneracy of the spectrum towards higher frequencies.
To compensate for this, we perform additional experiments where a weighting strategy is applied to the fingerprints, giving higher frequencies a lower weight for correlations.
Following this approach, we can restate the \gls{fcc} as
\begin{equation}
    \fccw \coloneq \frac{1}{\vert \bOmega \vert} \sum_{\bomega, \bomega' \in \bOmega} \left\vert r_\Theta(\bomega, \bomega') w(\bomega, \bomega') \right\vert
    \label{eq:fccw}
\end{equation}
introducing the weighting function $w(\bomega, \bomega')$ which depends on the frequency pair.
In the results depicted in~\autoref{fig:mse_fcc_fs_w_rx} we applied a linear weighting $w(\bomega, \bomega')\sim(\bomega+\bomega')^{-1}$.

In~\autoref{fig:fingerprints_1d_rx_w}, we can observe that, due to the weighting, correlations between coefficients with a higher frequency index are being reduced when directly comparing to~\autoref{fig:fingerprints_1d_rx}.
Furthermore, in~\autoref{fig:mse_fcc_fs_1d_rx_w} the \gls{fcc} of \emph{Circuit YZY} and \emph{Circuit 17} adjusts slightly, such that the relation to the \gls{mse} becomes more linear in general.
Identical experiments with Pauli-RY encoding showed a less prominent effect and are therefore omitted in this section.

\section{Existence of coefficient correlations}
\label{app:coeff_correlations}

While the existence of correlations between coefficient of the Fourier expansion of a \gls{qfm} is visible from the numerical experiments we provide in this work, the work from Ref.~\cite{nemkov_fourier_2023,wiedmann_fourier_2024} provides the foundation for a more formal proof.
Therefore we first restate a key finding from~\cite{wiedmann_fourier_2024}, \ie the formula for the exact coefficients dependent on $\btheta$:
\begin{equation}
    c_{\omega}(\btheta)=\sum_{\substack{s, c \in \mathbb{N}_{0}^{d} \\ s^{\prime}, c^{\prime} \in \mathbb{N}_{0}^{w}}} \frac{k_{s, c, s^{\prime}, c^{\prime}}(-\imag)^{\sum_{j=1}^{d} s_{j}}}{2^{\sum_{j=1}^{d}\left(s_{j}+c_{j}\right)}} \times p_{s, c}(\omega) \prod_{k=1}^{w} \sin \left(\theta_{k}\right)^{s_{k}^{\prime}} \cos \left(\theta_{k}\right)^{c_{k}^{\prime}}.
    \label{eq:exact_coefficients}
\end{equation}
Here $p(s, c, \omega)$ defines the sum of all coefficients belonging to $\omega$, with $s,c$ and $s^{\prime}, c^{\prime}$ counting the number of sine and cosine contributions by the input- and trainable parameters respectively.
The constant $k_{s, c, s^{\prime}, c^{\prime}}$ depends on the basis function following the notation in Ref.~\cite{wiedmann_fourier_2024}.

\textbf{Proposition}: Let $\theta \in[0, 2\pi)^{w}$ be fixed across $\omega$.
Then, even if $p(s, c, \omega)$ and $p\left(s, c, \omega^{\prime}\right)$ have disjoint supports, the functions $c_{\omega}(\btheta)$ and $c_{\omega^{\prime}}(\btheta)$ can exhibit nonzero correlation due to their dependence on the shared basis functions $\left\{\sin \left(\theta_{k}\right)^{s_{k}^{\prime}} \cos \left(\theta_{k}\right)^{c_{k}^{\prime}}\right\}$, \ie if we consider a single parameter $\theta_k \in \btheta$ per frequency $\omega$, the correlations between different $c_{\omega}(\theta)$ already exist due to the right-hand side of~\autoref{eq:exact_coefficients}.

The dependence on a shared basis function, however, does not imply that all \ase have equal distribution of correlations.
Depending on the circuit, $p(s, c, \omega)$ can vary, resulting in a \as specific \gls{fcc}.
As already stated in~\autoref{sec:method:correlations_in_qfm}, an \as with absolutely no correlation would be required to have an exponential number of parameters to account for the equally sized number of coefficients, rendering it untrainable in practice.

\section{Numerical Errors}
\label{app:num_errors}

Following the results of~\autoref{fig:mse_fcc_fs_1d_ry}, we provide a summary of the numerical errors in~\autoref{tab:meas_errors_fs_1d} as standard deviation $\sigma_m=\sqrt{\mathbb{E}\left[(\bar{m} - \mu(\bar{m}))^2\right]}$ of mean measurements $\bar{m}$ (expressibility, \gls{fcc} and \gls{mse}) for different \ase.
Similarly, we provide the errors for the 2D Fourier series (\cf \autoref{fig:mse_fcc_fs_2d}) and the \gls{hep} results (\cf \autoref{fig:mse_fcc_hep_2d}) in~\autoref{tab:meas_errors_fs_2d} and~\autoref{tab:meas_errors_hep_2d} respectively.
For all scenarios, we observe a low standard deviation consistent for all \ase and measurements.

\begin{table}[htbp]
    \centering
    \resizebox{\columnwidth}{!}{
        \begin{tabular}{lcccccccc}
            \toprule
            $\sigma$ & C 15          & C 16          & C 17          & C 18          & C 19          & C YZY         & C YZY Ent.    & \glsxtrshort{hea} \\
            \midrule
            Expr.    & \num{3.6e-03} & \num{5.4e-03} & \num{3.8e-03} & \num{3.8e-03} & \num{1.4e-03} & \num{6.6e-03} & \num{1.8e-03} & \num{1.8e-03}     \\
            FCC      & \num{9.6e-04} & \num{1.4e-03} & \num{1.6e-03} & \num{1.2e-03} & \num{9.3e-04} & \num{1.7e-03} & \num{1.2e-03} & \num{1.7e-03}     \\
            MSE      & \num{1.0e-03} & \num{2.4e-03} & \num{6.6e-05} & \num{4.4e-17} & \num{1.5e-03} & \num{3.3e-17} & \num{2.4e-03} & \num{1.5e-03}     \\
            \bottomrule
        \end{tabular}
    }
    \caption{Standard deviation of mean measurements(expressibility, \gls{fcc} and \gls{mse}) as depicted in \autoref{fig:mse_fcc_fs_1d_ry} across all seeds for different \ase.}
    \label{tab:meas_errors_fs_1d}
\end{table}

\begin{table}[htbp]
    \centering
    \resizebox{\columnwidth}{!}{
        \begin{tabular}{lcccccccc}
            \toprule
            $\sigma$ & C 15          & C 16          & C 17          & C 18          & C 19          & C YZY         & C YZY Ent.    & \glsxtrshort{hea} \\
            \midrule
            Expr.    & \num{3.6e-03} & \num{5.4e-03} & \num{3.8e-03} & \num{3.7e-03} & \num{1.4e-03} & \num{7.6e-03} & \num{2.8e-03} & \num{1.8e-03}     \\
            FCC      & \num{1.3e-04} & \num{2.9e-04} & \num{1.7e-04} & \num{2.4e-04} & \num{6.4e-04} & \num{2.5e-04} & \num{1.4e-04} & \num{1.6e-04}     \\
            MSE      & \num{2.9e-05} & \num{0.0e+00} & \num{2.2e-05} & \num{1.4e-19} & \num{3.5e-05} & \num{0.0e+00} & \num{4.6e-06} & \num{2.0e-05}     \\
            \bottomrule
        \end{tabular}
    }
    \caption{Standard deviation of mean measurements(expressibility, \gls{fcc} and \gls{mse}) as depicted in \autoref{fig:mse_fcc_fs_2d} across all seeds for different \ase.}
    \label{tab:meas_errors_fs_2d}
\end{table}

\begin{table}[htbp]
    \centering
    \resizebox{\columnwidth}{!}{
        \begin{tabular}{lcccccccc}
            \toprule
            $\sigma$ & C 15          & C 16          & C 17          & C 18          & C 19          & C YZY         & C YZY Ent.    & \glsxtrshort{hea} \\
            \midrule
            Expr.    & \num{3.6e-03} & \num{5.4e-03} & \num{3.8e-03} & \num{3.7e-03} & \num{1.4e-03} & \num{7.6e-03} & \num{2.8e-03} & \num{1.8e-03}     \\
            FCC      & \num{1.3e-04} & \num{2.9e-04} & \num{1.7e-04} & \num{2.4e-04} & \num{6.4e-04} & \num{2.5e-04} & \num{1.4e-04} & \num{1.6e-04}     \\
            MSE      & \num{2.9e-05} & \num{3.3e-05} & \num{3.0e-05} & \num{2.8e-05} & \num{2.4e-05} & \num{2.5e-05} & \num{3.0e-05} & \num{5.0e-05}     \\
            \bottomrule
        \end{tabular}
    }
    \caption{Standard deviation of mean measurements(expressibility, \gls{fcc} and \gls{mse}) as depicted in \autoref{fig:mse_fcc_hep_2d} across all seeds for different \ase.}
    \label{tab:meas_errors_hep_2d}
\end{table}

\section{Linking the FCC and the MSE}
\label{app:coeff_correlations_mse}

In this subsection, we derive how correlations between coefficients are reflected in the \gls{mse}.
For the sake of simplicity, we only consider one-dimensional inputs such that $\bx \mapsto x$ and $\bomega \mapsto \omega$, while $\btheta$ remains a vector of parameters.
Consider the target Fourier series $\hat{f}(x) = \sum_{\omega \in \Omega} \hat{c}_{\omega} \exp^{\imag\omega x}$ and the truncated Fourier series of the model $f(x, \btheta) = \sum_{\omega \in \Omega} c_{\omega}(\btheta) \exp^{\imag\omega x}$.
We now want to minimize a dataset sampled from $x\in \mathcal{X}$ using the \gls{mse}
\begin{equation}
    \begin{aligned}
        \loss(x, \btheta) & =\frac{1}{\sizex} \int_{x \in \mathcal{X}} \vert f(x, \btheta) - \hat{f}(x) \vert^2 dx                                                                        \\
                          & =\frac{1}{\sizex} \int_{x \in \mathcal{X}}\left|\sum_{\omega \in \Omega}\left(c_{\omega}(\btheta)-\hat{c}_{\omega}\right) \exp^{\imag \omega x}\right|^{2} dx \\
        \overset{(1)}     & {=}\sum_{\omega \in \Omega}\left|c_{\omega}(\btheta)-\hat{c}_{\omega}\right|^{2}                                                                              \\
    \end{aligned}
    \label{eq:loss_mse}
\end{equation}
where in (1) we assume the input interval $\mathcal{X}=[0, 2\pi)$ such that $\exp^{\imag\omega x}$ is orthonormal over $\mathcal{X}$ and apply Parseval's identity to simplify.
It should be noted that we assume the frequencies in the target function are equal to those of the \gls{qfm}, which is not the case in general but suffices for the study in this work.

As we want to argue about an arbitrary $\btheta \in \Theta$, but $\loss$ depends on a specific realization of $\btheta$, we must rewrite a statistical form of the loss $\exptheta[\loss(\btheta)]$ based on~\autoref{eq:loss_mse} as
\begin{equation}
    \begin{aligned}
        \loss(\btheta)                          & = \sum_{\omega}\left( \vert c_{\omega}(\btheta)\vert^{2} + \vert\hat{c}_{\omega}\vert^{2} - 2 \Re \left(c_{\omega}^*(\btheta) \hat{c}_{\omega} \right)\right)                                             \\
        \\
        \exptheta[\loss(\btheta)] \overset{(1)} & {=} \sum_{\omega \in \Omega}\left[\exptheta\left|c_{\omega}(\btheta)\right|^{2}+\left|\hat{c}_{\omega}\right|^{2}-2 \Re\left(\exptheta\left[c_{\omega}(\btheta)\right] \hat{c}_{\omega}^{*}\right)\right] \\
        \overset{(2)}                           & {=} \sum_{\omega \in \Omega}\left[\exptheta\left|c_{\omega}(\btheta)\right|^{2}+\left|\hat{c}_{\omega}\right|^{2}\right]                                                                                  \\
        \overset{(3)}                           & {=} \sum_{\omega \in \Omega}\left[\sigma_\Theta^2({c_{\omega}(\btheta)})+\left|\hat{c}_{\omega}\right|^{2}\right]                                                                                         \\
    \end{aligned}
    \label{eq:loss_mse_expanded}
\end{equation}
where in (1) we convert to a statistical loss with $\btheta \in \Theta$ and in (2) we assume zero mean based on~\autoref{app:variance}, \ie $\exptheta\left[c_{\omega}(\btheta)\right]=0$ such that $\Re\left(\exptheta\left[c_{\omega}(\btheta)\right] \hat{c}_{\omega}^{*}\right) = 0$ with $(\cdot)^*$ denoting the complex conjugate and $\Re(\cdot)$ the real part of a complex number.

Notably, $\loss$ is separable in $\omega$, therefore the final statement in~\autoref{eq:loss_mse_expanded} only captures the standard deviation of each coefficient individually, corresponding to the diagonal of the covariance matrix $K$.
The correlation in the off-diagonal elements, however, puts additional constraints on the optimizer, as individual coefficients can't be adjusted individually.
We can capture this in $\mathcal L$ as well, by assuming that each coefficient is actually not dependent on a single frequency, but on all other frequencies in $\Omega$ instead, as discussed in~\autoref{app:coeff_correlations}.
This can be accounted for by substituting $c_\omega(\btheta) = \sum_{\omega' \in \Omega} \zeta(\omega, \omega'; \btheta)$, where $\zeta(\omega, \omega'; \btheta)$ is a proxy coefficient, that shares contributions to $\omega$ and $\omega'$ given the parameter vector $\btheta$.
The proxy coefficient for $\omega = \omega'$ would then reflect the prior assumption of a fully separable loss, still allowing for a scenario where the coefficients are fully independent, \ie where $\zeta(\omega, \omega'; \btheta) = 0 \ \forall \ \omega \neq \omega'$ such that $c_\omega(\btheta) = \zeta(\omega, \omega; \btheta)$.

Given this assumption, we restate the second last equation~\autoref{eq:loss_mse_expanded} as follows:
\begin{equation}
    \begin{aligned}
        \exptheta[\loss(\btheta)] & = \sum_{\omega \in \Omega}\left[\exptheta\left|\sum_{\omega' \in \Omega} \zeta(\omega, \omega'; \btheta)\right|^{2}+\left|\hat{c}_{\omega}\right|^{2}\right]                                                           \\
        \overset{(1)}             & {=} \sum_{\omega \in \Omega}\left[\sum_{\omega', \omega'' \in \Omega} \left[\exptheta\left[ \zeta(\omega, \omega'; \btheta) \zeta^*(\omega, \omega''; \btheta) \right]\right]+\left|\hat{c}_{\omega}\right|^{2}\right] \\
        \overset{(2)}             & {=} \sum_{\omega \in \Omega}\left[\sum_{\omega', \omega'' \in \Omega} r_\Theta(c_{\omega'}, c_{\omega''}) \sigma_\Theta({c_{\omega'}}) \sigma_\Theta({c_{\omega''}}) + \left|\hat{c}_{\omega}\right|^{2}\right]        \\
    \end{aligned}
    \label{eq:loss_mse_expanded_fcc}
\end{equation}
where in (1) we expanded the complex inner product that requires introducing two additional dummy frequencies $\omega'$ and $\omega''$, which can be regarded as directional interactions between frequencies.

From~\autoref{eq:pearson_correlation} we restate the Pearson correlation between coefficients of frequency $\omega$ and $\omega'$ in a slightly different form as
\begin{equation}
    r_\Theta(c_\omega, c_{\omega'}) = \frac{K_\Theta(c_{\omega}, c_{\omega'})}{\sqrt{\sigma_\Theta^2({c_{\omega}}) \sigma_\Theta^2({c_{\omega'}})}} \overset{\bar c = 0}{=} \frac{\exptheta\left|c_{\omega}(\btheta)c_{\omega'}(\btheta)\right|}{\sigma_\Theta({c_{\omega}}) \sigma_\Theta({c_{\omega'}})}
    \label{eq:correlation}
\end{equation}
with $\exptheta(\cdot)$ being the expectation value of $(\cdot)$ over $\Theta$, $\sigma_\Omega^2({c_{\omega}})=\text{Var}_\Omega(c_{\omega})=\exptheta\left|c_{\omega}(\btheta)c_{\omega'}(\btheta)\right|$ being the standard deviation of $c_{\omega}$, $K_\Theta(c_{\omega}, c_{\omega'})$ the covariance between $c_{\omega}$ and $c_{\omega'}$ and $\bar c_{\omega}$ the mean over $\Theta$.
Using the right-hand side of~\autoref{eq:correlation} we can substitute the inner term in (2), which introduces the Pearson correlation to the overall loss function.

As stated in~\autoref{eq:loss_mse_expanded_fcc}, the \gls{fcc} is the sum of over all frequency pairs of $r$ and therefore linearly related.
Therefore, the \gls{fcc} is also closely related to $\loss$ but strongly weighted by the variances of the coefficients.
Following the weighting strategy introduced in~\autoref{app:weighting_the_fcc}, we can set $w(\omega', \omega'')=\sigma_\Theta({c_{\omega'}}) \sigma_\Theta({c_{\omega''}})$ and therefore forcefully restating the final loss equation as follows:
\begin{equation}
    \begin{aligned}
        \exptheta[\loss(\btheta)] & = \sum_{\omega \in \Omega}\left[\sum_{\omega', \omega'' \in \Omega} r_\Theta(c_{\omega'}, c_{\omega''}) w(\omega', \omega'') + \left|\hat{c}_{\omega}\right|^{2}\right] \\
                                  & = \sum_{\omega \in \Omega}\left[\fccw\sizefreq^{-1} + \left|\hat{c}_{\omega}\right|^{2}\right]                                                                          \\
                                  & = \fccw \sum_{\omega \in \Omega}\left|\hat{c}_{\omega}\right|^{2}.
    \end{aligned}
\end{equation}
Here, only the \gls{fcc} and the contribution of the squared absolute value of each target coefficient remain.
The findings from Ref.~\cite{mhiri_constrained_2024} suggest an exponential decay of the variance towards high frequency components, providing a practical hint on estimating $\sigma_\Theta$.
This equation analytically summarizes our main contribution of this work; the loss of learning a \gls{qfm} on a Fourier series with the same number of frequencies is linearly dependent on the (weighted) \gls{fcc}, \ie the correlation between the frequency components of the \gls{qfm}.

\section{Computing the expressibility \& complexity estimate}
\label{app:computing_the_complexity_estimate}

The expressibility is a commonly used metric that quantifies how a given \as can cover the Hilbert space spanned by a \gls{pqc}.
It can be calculated using the KL-divergence~\cite{kullback_information_1951} between the model $\int_{\btheta}\left(\left|\psi_{\btheta}\right\rangle\left\langle\psi_{\btheta}\right|\right)^{\otimes t} d \btheta$ and the distribution of the Haar-integral $\int_{\text {Haar }}(|\psi\rangle\langle\psi|)^{\otimes t} d \psi$ of a state $t$-design with $D_{\mathrm{KL}}\left(\hat{P}_{\text {Model }}(F ; \btheta) \| P_{\text {Haar }}(F)\right)$.
Here $F=\left|\left\langle\psi_{\boldsymbol{\varphi}} \mid \psi_{\boldsymbol{\phi}}\right\rangle\right|^2$ describes the fidelity between two states.

Generally speaking, an ansatz with a high expressibility is expected to have a high chance of reaching an arbitrary state, as parameters are likely to be sampled in the desired area of the Hilbert space.
For this reason, the expressibility can be used as a metric to estimate the performance of a given \as.
However, it should also be noted that \ase with high expressibility are also more prone to the Barren Plateau phenomenon as described in Ref.~\cite{ragone_lie_2024}.

Calculating the expressibility is a computationally expensive task as it requires simulation of the density matrix.
While the distribution of the Haar integral can be cached, the distribution of the model requires sampling the parameters and therefore varies with the number of layers.
Therefore, the overall complexity is at $\mathcal{O}(M N^3)$ (matrix-matrix multiplication) where $M$ is the number of parameter samples used to calculate the expressibility and $N=2^n$ is the dimensionality of the Hilbert space spanned by $n$ qubits.
Furthermore, the expressibility can only be calculated in simulation, as access to the density matrix is naturally not available on real hardware.

In comparison, the \gls{fcc} metric only requires computing the expectation value (matrix-vector multiplication) but adds the overhead of the \gls{fft} $\mathcal{O}(K\log(K))$ where $K$ is the number of input samples, resulting in a complexity of the \gls{fcc} being $\mathcal{O}(M N^2 K\log(K))$.
To prevent aliasing, $K$ has to be $\ge \nscrit$, where $\hnscrit$ is the highest frequency of the \gls{qfm}.
For the case of a $D$-dimensional input, the number of samples required is $K \ge (\nscrit)^D$.
Therefore, the overall complexity for computing the \gls{fcc} of a $n$-qubit \gls{qfm} with $D$-dimensional input and the highest available frequency $\hnscrit$ becomes $\mathcal{O}(M N^2 (\hnscrit)^D\log(\hnscrit))$, when just satisfying the sampling theorem.
Considering a window length $K$ being power of $2$, the complexity of the \gls{fft} can further be reduced according to Ref.~\cite{cooley_algorithm_1965} to $K/2\log_2(K)$ which, however, does not reduce the overall complexity in the asymptotic limit.

In case of having just a single-layered \gls{qfm}, where the highest frequency is just $n$~\cite{schuld_effect_2021}, it allows us to simplify the expression to $\mathcal{O}(M N^2 n^D \log_2(n))$.
In a scenario where the number of qubits is equal to the number of layers, we find that $\hnscrit=n^2$ and the \gls{fcc} complexity becomes $\mathcal{O}(M N^2 n^{2D} \log_2(n))$ which is still easier to compute than the expressibility.
The worst case (and unlikely) scenario would be a single qubit \gls{qfm} with $n$ layers, where the (theoretically) highest frequency is still $n$, but $N=2$ and therefore the computational complexity of the expressibility would reduce to $\mathcal{O}(M)$, while the \gls{fcc} is $\mathcal{O}(M n^D \log_2(n))$.
To summarize, a comparison with the expressibility complexity $\mathcal{O}(M 2^{3n})$ shows that in the limit of $n$, the \gls{fcc} with $\mathcal{O}(M N^2 (\hnscrit)^D\log(\hnscrit))$ is easier to compute, regardless of $D$ and any (feasible) constraint on the number of layers.

Furthermore, the \gls{fcc} metric is also more suitable for real hardware as the expectation value can be directly calculated on the device.
In contrast, the expressibility requires full-state tomography which is exponentially costly in the number of qubits, whereas the \gls{fcc} requires only a single measurement basis.
The error introduced by shot noise is $\nicefrac{1}{\sqrt{N_{\text{shots}}}}$ with $N_{\text{shots}}$ being the number of shots taken and applicable to both measures.

\section{High-Energy Physics}
\label{app:high-energy-physics}

\paragraph*{Additional numerical results}

\begin{figure}[htb]
    \begin{subfigure}[c]{0.42\textwidth}
        \includegraphics[clip, trim=0.0cm 0.5cm 0.5cm 2.0cm, width=\textwidth]{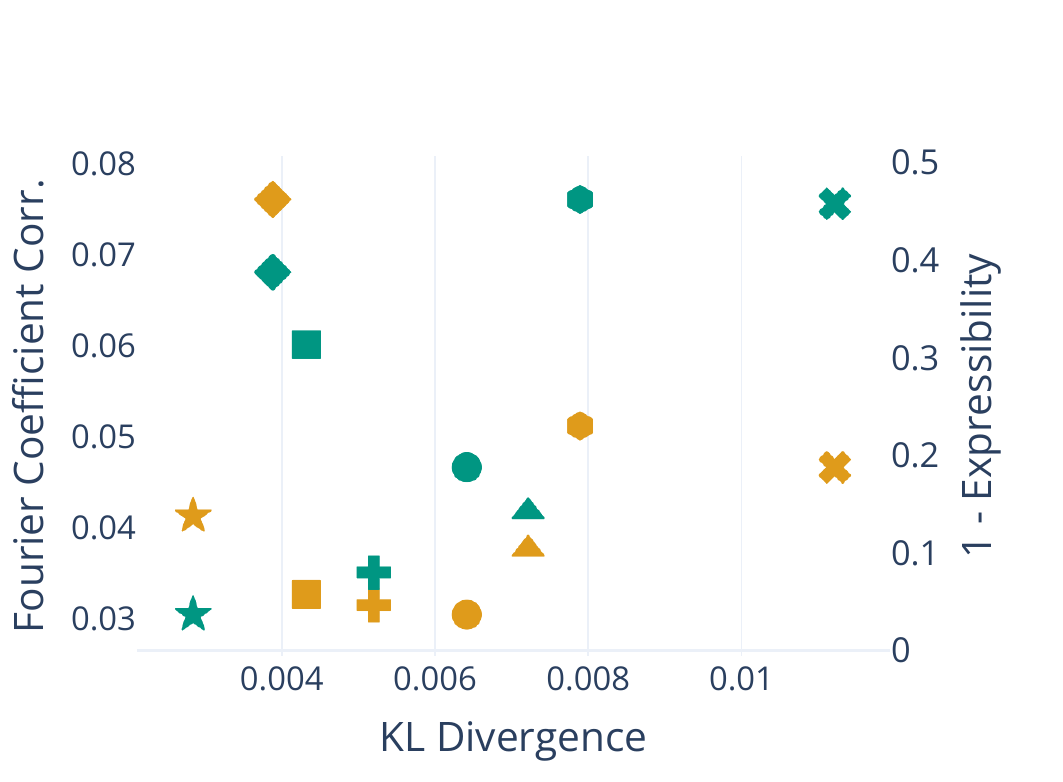}
        \caption{}
        \label{fig:kl_divergence_fcc_hep_2d}
    \end{subfigure}
    \begin{subfigure}[c]{0.42\textwidth}
        \includegraphics[clip, trim=0.0cm 0.5cm 0.5cm 2.0cm, width=\textwidth]{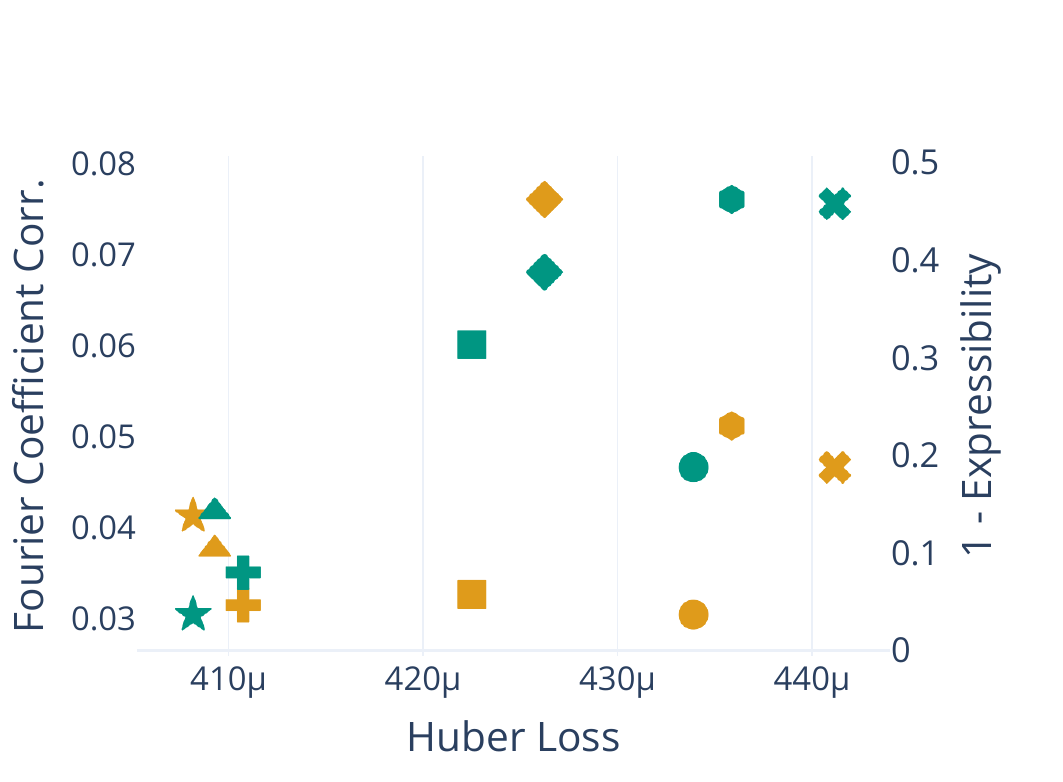}
        \caption{}
        \label{fig:huber_loss_fcc_hep_2d}
    \end{subfigure}
    \begin{subfigure}[c]{0.14\textwidth}
        \includegraphics[clip, trim=2cm 0cm 10.7cm 2.5cm, width=\textwidth]{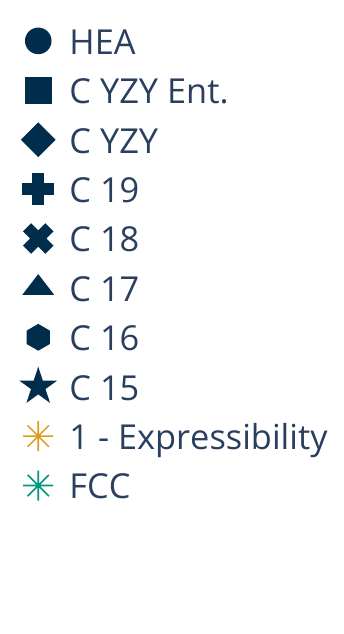}
    \end{subfigure}
    \caption{Comparison of \textcolor{color1}{FCC} and \textcolor{color2}{expressibility complement} over the \gls{kl} divergence (a) and Huber loss (b) for different \ase (color coded). Training on 2D \gls{hep} dataset with averaged over \num{10} independent dataset seeds and \num{10} distinct parameter initializations.}
    \label{fig:add_metrics_fcc_hep_2d}
\end{figure}

In the following, we provide additional numerical results for training on the \gls{hep} dataset where the setup is identical to~\autoref{sec:results:hep} but, instead of the \gls{mse}, compare the \gls{kl} divergence and Huber loss to the \gls{fcc} and expressibility.
The result of this comparison can be seen in~\autoref{fig:kl_divergence_fcc_hep_2d} and~\autoref{fig:huber_loss_fcc_hep_2d} respectively.

For the \gls{kl}-divergence, the \gls{fcc} and expressibility both fail as predictors, although \emph{Circuit 15} and \emph{Circuit 18} still represent the best and worst performing circuits, as already seen in \autoref{fig:hep_results}.
In contrast, the results for the Huber loss are almost identical with the results for the \gls{mse} as shown in~\autoref{fig:hep_results}.

\paragraph*{Physical foundations}

Next, we want to provide an explanation on the physical foundations of the \gls{hep} dataset used in this work.
In high-energy proton–proton collisions at the \gls{lhc}, the dominant processes involve interactions between constituent partons; quarks and gluons.
These quark–gluon scatterings lead to the production of jets, collimated sprays of hadrons that are experimentally observed as signatures of the underlying partonic activity.

\begin{figure}[htb]
    \centering
    \begin{minipage}[c]{0.2\textwidth}
        \centering
        \scalebox{0.6}{
            \begin{tikzpicture}
                \begin{feynman}
                    \setlength{\unitlength}{.75cm}\large
                    \vertex (q_in) at (1,7) {$q$};
                    \vertex (g_in) at (1,2) {$g$};
                    \vertex (int1) at (3,4);
                    \vertex (int2) at (3,5);
                    \vertex (q_out) at (5,7) {$q$};
                    \vertex (g_out) at (5,2) {$g$};
                    \diagram* {
                    (q_in) -- [fermion] (int2) -- [fermion] (q_out),
                    (g_in) -- [gluon] (int1) -- [gluon] (g_out),
                    (int1) -- [gluon] (int2)
                    };
                \end{feynman}
            \end{tikzpicture}
        }
    \end{minipage}
    \begin{minipage}[c]{0.2\textwidth}
        \centering
        \scalebox{0.6}{
            \begin{tikzpicture}
                \begin{feynman}
                    \setlength{\unitlength}{.75cm}\large
                    \vertex (q_in) at (1,7) {$q$};
                    \vertex (g_in) at (1,2) {$g$};
                    \vertex (int1) at (3,4);
                    \vertex (int2) at (3,5);
                    \vertex (q_mid) at (4,6);
                    \vertex (q_out) at (5,7) {$q$};
                    \vertex (g_out) at (5,2) {$g$};
                    \vertex (g_emit) at (5,5) {$g$};
                    \diagram* {
                    (q_in) -- [fermion] (int2) -- [fermion] (q_mid) -- [fermion] (q_out),
                    (g_in) -- [gluon] (int1) -- [gluon] (g_out),
                    (int1) -- [gluon] (int2),
                    (q_mid) -- [gluon] (g_emit)
                    };
                \end{feynman}
            \end{tikzpicture}
        }
    \end{minipage}\hfill
    \begin{minipage}[c]{0.5\textwidth}
        \caption{Quark–gluon scattering at \gls{lo} and \gls{nlo}. The additional gluon emission in \gls{nlo} corresponds to an extra vertex with a factor of the strong coupling constant $\alpha_s$, reducing the interaction probability and increasing the number of jets observed.}
        \label{fig:qg_two_diagrams}
    \end{minipage}

\end{figure}
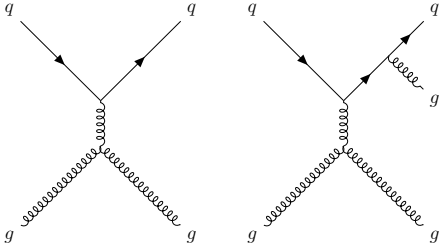

At \gls{lo}, quark–gluon scattering produces two outgoing partons that hadronize into two jets, corresponding to the final-state quark and gluon, see~\autoref{fig:qg_two_diagrams} left.
The corresponding Feynman diagram features a single gluon exchange vertex.

Beyond \gls{lo}, at \gls{nlo} and higher, additional gluon emissions occur, see~\autoref{fig:qg_two_diagrams} right.
Each emission corresponds to an extra vertex in the Feynman diagrams, contributing a factor of the strong coupling constant $\alpha_s$, which quantifies the interaction strength of \gls{qcd}.
Since $\alpha_s$ runs with the energy scale, it decreases logarithmically at higher momentum transfers (asymptotic freedom).
This running controls the probability of multiple gluon emissions and thus the jet multiplicity.

The presence of three or more jets arises naturally from these additional gluon emissions during the parton shower phase, reflecting the increased complexity of the event.
The number of jets and their momentum distribution provide direct insight into the \gls{qcd} dynamics and the value of $\alpha_s$ at the relevant scale.

The center-of-mass energy $E_{\mathrm{cm}}$ plays a central role because it sets the scale for the hard scattering process.
Higher $E_{\mathrm{cm}}$ implies access to larger momentum transfers, reducing $\alpha_s$ and opening greater phase space for gluon radiation.
Consequently, at \gls{lhc} energies, this results in higher jet multiplicities and broader momentum spectra, making $E_{\mathrm{cm}}$ a fundamental metric for studying jet formation and modeling \gls{qcd} interactions.

\section{Sampling}
\label{app:sampling}

\paragraph*{Shot noise}

Considering the expectation value of a \gls{qfm} $f(\bx, \btheta)  =\bran{0} U^{\dagger}(\bx, \btheta) \mathcal{M} U(\bx, \btheta)\ketn{0}$, it is commonly known that the true error due to a finite number of $S$ shots is given by $\nicefrac{1}{\sqrt{S}}$.
Therefore, we can state $\tilde f$, which is the expectation value $f$ with some shot noise $\delta(\bx)$.
\begin{equation}
    \tilde{f}(\bx, \btheta) = f(\bx, \btheta) + \delta(\bx) \quad \text{with} \quad \vert \delta(\bx) \vert \leq \frac{1}{\sqrt{S}} \ \forall \ \bx
\end{equation}

Due to the linearity of the \gls{fft} and $\Delta c_{\bomega} = \tilde{c}_{\bomega} - c_{\bomega}$ being the deviation in the Fourier coefficients caused by shot noise and we can say that $\Delta c_{\bomega} = \sum_{\bx} \delta(\bx) \exp{(\imag \bomega \bx)}$.
With $\delta(\bx) \leq \frac{1}{\sqrt{S}}$, we have that
\begin{equation}
    \vert \Delta c_{\bomega} \vert \leq \sum_{\bx} \vert \delta(\bx) \vert \cdot \vert e^{\imag \bomega \bx} \vert = \sum_{\bx} \vert \delta(\bx) \vert \leq \frac{1}{\sqrt{S}} \sizex.
\end{equation}
Because for a single \gls{fft}, we require $\sizex$ sampling points, the error for a single frequency coefficient accumulates linearly while the number of shots reduces this error by $\frac{1}{\sqrt{S}}$.

\paragraph*{Sampling error}

The Pearson correlation method requires statistical information of $c_{\bomega}(\btheta)$ over $\Theta$.
For any practical scenario, the actual number of samples $M$ drawn from $\Theta$ is therefore important to consider.
We assume that $c_{\bomega}(\btheta)$ is zero-centred (\cf~\autoref{app:variance}) and because $f \in [-1,1]$, the energy of the Fourier transform is bounded, requiring the coefficients to be bounded as well.
The standard error for $r$ given a finite sample size $M$ is given by~\cite{bowley_standard_1928}
\begin{equation}
    \sigma_{r_{\bomega, \bomega'}} \approx \sqrt{\frac{1-r_{\bomega, \bomega'}^2}{(M-1)^2}},
\end{equation}
where $\sigma_{r_{\bomega, \bomega'}}$ denotes the deviation from the true Pearson correlation $r$ that converges to $0$ in the limit of $M \to \infty$.
As the \gls{fcc} is just the summation of $r_{\bomega, \bomega'}$ over all frequency pairs, we can estimate a similar error as
\begin{equation}
    \sigma_{FCC} \approx \frac{1}{\sizefreq} \sum_{\bomega, \bomega'} \sqrt{\frac{1-r_{\bomega, \bomega'}^2}{(M-1)^2}} \overset{(1)}{=} \frac{\sqrt{1-\bar{r}^2}}{\sizefreq(M-1)}
\end{equation}
where, in (1), we assume independence of the correlation estimates across frequency pairs, \ie all correlations have a similar variance.
Therefore, we can utilize the average correlation $\bar{r}$ of a \gls{qfm} and the error in the \gls{fcc} scales as $\sim\frac{1}{M}$ with the number of samples $M$.

\paragraph*{Variances of coefficients}

Practically, we already know that the variance of $c_{\bomega}$ decays exponentially with the frequency (\cf~\autoref{app:variance}), which we can model by $\sigma^2_\Theta(c_{\bomega}) \sim \beta e^{-\alpha\|\boldsymbol{\bomega}\|}$ with $\beta>0$ and $\alpha>0$ being some constants depending on the specific \gls{fm} and \as.
Because of this exponential decay in the denominator of $r_{\bomega, \bomega'}$, $\sigma_{FCC}$ is dominated by the sampling variability for higher frequencies.
We can roughly estimate this by $\sigma_{FCC} \approx \frac{1}{M} e^{\alpha(\Vert\bomega\Vert + \Vert\bomega'\Vert)}$
This result implies an exponential requirement on $M$ already, if high frequency accuracy is required.
However, if we consider that redundancies decay towards higher frequencies as well~\cite{mhiri_constrained_2024}, we can expect fewer parameters contributing to high-frequent $c_{\bomega}$, which even increases the requirement on the number of parameter samples $M$.

\paragraph*{Number of qubits}

By now, we considered a fixed number of qubits, and therefore also a fixed length of the parameter vector $\vert \btheta \vert$ (given a particular \as).
Changing the number of qubits naturally changes the number of parameters and, therefore, the requirement on $M$.
However, because the Hilbert space increases with $2^n$, the question arises whether there is an even stronger requirement on $M$ than just compensating for the higher number of parameters.
To show that $M$ does not have to scale exponentially with $n$, we consider an average expectation value $\bar{f}$ as
\begin{equation}
    \bar{f}(\bx) = \frac{1}{M} \sum_{\btheta \in \Theta} f(\bx, \btheta) \quad \text{with} \quad \sizeparams = M.
\end{equation}
Because the expectation value is bounded in $[-1,1]$, $\bar{f}$ must be bounded as well, and we can apply Hoeffding's inequality
\begin{equation}
    \mathbb{P}\left(\left\vert \hat{\bar{f}} - \bar{f} \right\vert \geq \epsilon \right) \leq 2 \exp\left(-2M\epsilon^2\right)
\end{equation}
which allows us to estimate the probability that an average $\hat{\bar{f}}$ deviates from the true $\bar{f}$ by less than $\epsilon$.

\end{document}